\documentclass[%
 reprint,
superscriptaddress,
 aps,pre,
]{revtex4-1}
\usepackage{amsmath} 
\usepackage{color}
\usepackage{dcolumn}
\usepackage{bm}

\usepackage{color}

\makeatletter
\newcommand{\overleftrightsmallarrow}{\mathpalette{\overarrowsmall@\leftrightarrowfill@}}
\newcommand{\overrightsmallarrow}{\mathpalette{\overarrowsmall@\rightarrowfill@}}
\newcommand{\overleftsmallarrow}{\mathpalette{\overarrowsmall@\leftarrowfill@}}
\newcommand{\overarrowsmall@}[3]{%
  \vbox{%
    \ialign{%
      ##\crcr
      #1{\smaller@style{#2}}\crcr
      \noalign{\nointerlineskip}%
      $\m@th\hfil#2#3\hfil$\crcr
    }%
  }%
}
\def\smaller@style#1{%
  \ifx#1\displaystyle\scriptstyle\else
    \ifx#1\textstyle\scriptstyle\else
      \scriptscriptstyle
    \fi
  \fi
}
\makeatother

\newcommand{\te}[1]{\overleftrightsmallarrow{#1}}

\usepackage{amssymb} 
\usepackage{wrapfig}
\usepackage{graphicx}
\usepackage{amsmath}
\usepackage{textcomp}
\usepackage{placeins}

\usepackage{textcomp}

\begin{document}


\title{Impedance resonance in narrow confinement}

\author{Sonja Babel}
\email{sonja.babel@uni-duesseldorf.de}
\affiliation{Institut f\"ur Theoretische Physik II: Weiche Materie, Heinrich-Heine-Universit\"at D\"usseldorf, Universit\"atsstra\ss e 1, D-40225 D\"usseldorf, Germany}

\author{Michael Eikerling}
\affiliation{Department of Chemistry, Simon Fraser University, 8888 University Drive, Burnaby, British Columbia, Canada V5A 1S6}

\author{Hartmut L\"owen}
\affiliation{Institut f\"ur Theoretische Physik II: Weiche Materie, Heinrich-Heine-Universit\"at D\"usseldorf, Universit\"atsstra\ss e 1, D-40225 D\"usseldorf, Germany}


%


\begin{abstract}
  The article explores the ion flux response of a capacitor configuration to an alternating voltage. The model system comprises a symmetric binary electrolyte confined between plan-parallel capacitor plates. The AC response is investigated for the sparsely studied albeit practically important case of a large amplitude voltage applied across a narrow device, with the distance between the two plates amounting to a few ion diameters. Dynamic density functional theory is employed to solve for the spatiotemporal ion density distribution as well as the transient ion flux and complex impedance of the system. The analysis of these properties reveals a hitherto hidden impedance resonance. A single ion analogue of the capacitor, which is equivalent to neglecting all interactions between the ions, is employed for a physical interpretation of this phenomenon. It explains the resonance as a consequence of field-induced ion condensation at the capacitor plates and coherent motion of condensed ions in response to the field variation.   
\end{abstract}
\maketitle
\section{Introduction}
There is an escalating interest in understanding capacitive phenomena in narrowly confined ionic systems \cite{Jiang2011, Kornyshev2013, Rochester2016, Kong2015}. Obviously, the topic is of foremost fundamental importance in electrochemistry. Besides, it is of practical relevance for analyzing the dynamic response of charged colloidal systems \cite{Leunissen2005,Demirors2015,Wurger2008} as well as nanoelectrochemical systems \cite{Arico2005,Montelongo2017} to a varying electric field, as encountered for instance in electroactuators \cite{Baughman1999,Kornyshev2017} or capacitive deionization systems \cite{Porada2013,Biesheuvel2010,Haertel2015JPCM}. The size of ions in relation to the size of the confining systems is a crucial consideration in such systems. 
Specifically, for ionic systems confined to lengths of the order of the ion diameter steric effects become important.

The archetypal capacitive system consists of a liquid electrolyte or ionic liquid that is confined by rigid walls made of metallic conductor insulated against the electrolyte, as depicted in fig.~\ref{fig_systemsketch}. The model system used here is infinite in lateral direction, rendering the problem effectively one-dimensional. Note that there is thus no bulk from which ions are taken or to which ions can leave nor are charges transfered from the ions to the walls. Ions only move back and forth within the gap. The primordial scientific interest lies in understanding the response of such a system to a modulation of the applied metal-phase potential. The response function in question is the result of a complex interplay between variations in metal surface charge density, electrolyte potential and ion density distribution. In this work, classical {\it dynamic} density functional theory (DDFT) \cite{Evans2016,Archer2004,Espanol2009,Marconi1999} is used to study the ion dynamics in a narrowly confined electrolyte slab, whose thickness equals a few ionic diameters. The ionic system is exposed to a dynamic voltage between the capacitor plates with harmonic (sinusoidal) time dependence determined by the angular frequency $\omega$ and the amplitude $\Delta U$.

The motivation to study this model is threefold: 
first of all, the model is best applied to the mesoscopic scale
for oppositely charged colloids \cite{Leunissen2005,Demirors2015}.
Using organic solvents these can be prepared even at low concentrations of dissolved ionic countercharges such that the microion concentration is small \cite{Royall}. These dispersions have been exposed to DC \cite{lanes}
and AC \cite{bands} electric fields that gave rise to strong spatio-temporal responses.
The presented model system resembles the configuration considered in Ref.\citenum{bands} though the width of the electrolyte-filled slit between insulating walls in our case corresponds to a few colloidal layers only. Apparent issues in applying our model to colloids lie in ignoring the residual microion concentration, which however can be kept to small (micromolar) concentrations in organic solvents, and neglecting the hydrodynamic interactions mediated by the solvent, which can be justified by using particles whose hydrodynamic radius is much smaller than their interaction radius, as is the case for solvent permeable particles \cite{Saunders19991, Holmqvist2012}. The mentioned problems are also mitigated by the separation in the frequency domain of the responses of microions and solvent molecules on the one and the colloidal ions on the other hand. The colloidal system can of course be scaled down in size towards oppositely charged micelles and nanocolloids.

Secondly, one can think about a molecular realization in nanogap electrodes
with the capacitor walls electrically isolated from the electrolyte in order to prevent any ion oxidation at the walls. The peculiar geometry
has not yet been exploited experimentally but is in principle feasible \cite{Lemay2013}. It should be noted though that applying our model to this case implies neglecting any specific solvent contributions to the AC electric field response as well as any molecular and specific details of the walls \cite{Biesheuvel2018}.

Thirdly, the considered model logically extends basic principal model studies of capacitor configurations that can be traced back to classical works of Gouy (1910) \cite{Gouy1910} and Chapman (1913) \cite{Chapman1913}. Form these early works, basic problems of electrified interfaces and confined electrolytes were approached using continuum theories based on Poisson-Boltz\-mann (PB) and Poisson-Nernst-Planck (PNP) equations. Later on these classical continuum theories 
were modified to account for steric effects induced by finite ion size \cite{Bikerman1942, Borukhov1997, Kornyshev2007, Burger2012, Siddiqua2017, Meng2014,Qiao2016,Jiang2014,Lian2016} and specific solvent polarization effects \cite{Gongadze2015}.

Our model system is similar to those considered in recent theoretical studies by Beunis et al.~\cite{Beunis2008}, Olesen et al.~\cite{Bazant2010} and Feicht et al.~\cite{Feicht2016}. However, in contrast to those works, we consider a case of more narrow ion confinement, wherein the width of the electrolyte slab, $L$, is of the order of several ion diameters, $\sigma$, i.e., $\sigma \le L \le 4 \sigma$, and we focus entirely on the limit of large electric fields. 
We employ a classical {\it dynamic} density functional theory \cite{Evans2016,Archer2004,Espanol2009,Marconi1999} (DDFT), explicitly including the steric repulsion between the ions. A similar model has been used before in Ref.\citenum{Jiang2014,Lian2016} to consider the charging kinetics of an electric double layer in response to a voltage step.

Here we apply this approach to study the capacitive response of the ionic system to a transient electric field. Our system is smaller than the one that was considered in Refs.\citenum{Jiang2014,Lian2016} and we are interested in the ion flux response to an AC voltage signal with large amplitude. This response function should be amenable to experimental study using electrochemical impedance spectroscopy (EIS).

DDFT is known to be computationally highly efficient and it allows geometric parameters like ion diameter and slab thickness
to be widely tuned. With DDFT the system response can be studied under large amplitude $\Delta U \gg U_{\rm T}$, with the thermal voltage $U_{\rm T} = k_{\rm B} T/q$, where $k_{\rm B}$ is the Boltzmann constant, $T$ the temperature, and $q$ the ion charge, and over a wide range of $\omega$. It is thus an ideally suited tool to explore ion dynamics in a capacitor configuration in the limit of strong ion confinement \cite{Jiang2011, Kornyshev2013, Rochester2016, Kong2015} where the full interplay of steric correlation effects, electrostatic interactions, as well as ion transport by diffusion and migration unfolds.

In the following section, we introduce the model system and describe the physical-computational methodology based on dynamic density functional theory. Equations are non-dimensionalized and typical parameter sets are discussed. In the results section, we analyze and discuss the dynamic density profiles of ions and the impedance response of the microscopic model system. A single-ion capacitor model is presented to explain the observed resonance effect in the impedance.

\section{Model}
\begin{figure}[htb]
\hspace*{1cm}
\includegraphics[width=0.6\columnwidth]{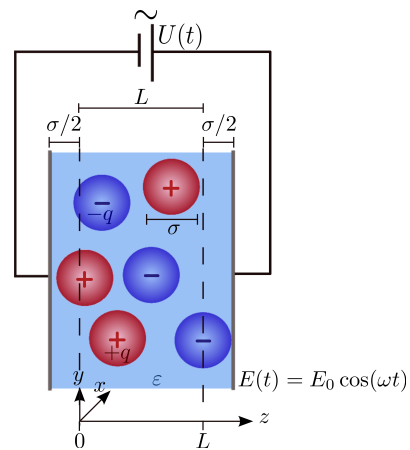}
\caption{Sketch of the system. Balanced numbers of equally sized ions with diameter $\sigma$ and charge $q_\pm=\pm q$, are kept in a slab between two plan-parallel insulating capacitor plates in an electrolyte of relative dielectric constant $\varepsilon$. The width $L$ denotes the accessible length in $z$ direction. For simplicity the system is assumed to be infinitely extended in $xy$ direction. An alternating external voltage $U(t)$ with amplitude $\Delta U$ produces an oscillating electric field $E(t)$ acting on the ions.}\label{fig_systemsketch}
\end{figure}

An electro-neutral mixture of colloidal cations and anions with equal charge magnitude $q$ and equal diameter $\sigma$ is kept in a stagnant fluid  with dielectric constant $\varepsilon$. This ionic system \cite{Levin,Fisher1993,vanRoij1997,Gillespie2001,Kornyshev1995,Kornyshev2007,Kornyshev2014,Messina2000, Lobaskin2007,Fennell2009, Girotto2017} is confined between two infinitely extended plan-parallel capacitor plates \cite{Bazant2004,Beunis2008,Feicht2016,Bazant2009,Bazant2010,Haertel2015PRE, Haertel2016,HaertelReview2017,Gillespie2015,Haertel2015EnEnvSci}, see fig.~\ref{fig_systemsketch}. The plates are polarized with an external alternating voltage $U(t)$ that creates an oscillating electric field $E(t)$ across the electrolyte slab. The ions are modeled within the restricted primitive model \cite{Orkoulas1999, Luijten2002, Yan1999, Hynnen2006, Panagiotopoulos2002, Caillol2002, Valleau1998, Carvalho1995, Esselink1996, Alts1987} as hard charged spheres of diameter $\sigma$ interacting via steric and Coulomb interactions. The capacitor plates are introduced as hard insulating walls leading to a no-flow condition for the ions. The outermost possible positions of the ion centres are then situated at a distance $\sigma/2$ away from the physical walls. $L$ denotes the accessible width perpendicular to the capacitor plates. The system is considered in the highly confined limit where $L$ is of the order of $\sigma$.

We solve the model for the time- and space-dependent densities of the two ion species using DDFT \cite{Evans2016,Archer2004,Espanol2009,Marconi1999}, which is the time-dependent variant of classical DFT \cite{HansenLoewen2000,Evans1979,Gillespie2002,Gillespie2003,Gillespie2005,Gillespie2011,Roth2005,Roth2016, Patra1994, Patra1994n2,Patra1997,Patra1999,Kierlik1991}. From the densities we also obtain the charge flux in the system. The linear response part of the current is used to further calculate a quantity that is analogous to a local impedance of the capacitor configuration. Subsequently, we will therefore refer to it as the local impedance.

\subsection{Theory}
\subsubsection{Dynamic density functional theory (DDFT)}
Dynamic density functional theory relates the time evolution of the density to the functional derivative of the free energy of the system in the form 
\begin{equation}
\frac{\partial \rho_\pm\left(\vec{r},t\right)}{\partial t}=\beta D\nabla\left[\rho_\pm\left(\vec{r},t\right)\nabla\left(\frac{\delta\mathcal{F}\left[\rho_+\left(\vec{r},t\right),\rho_-\left(\vec{r},t\right)\right]}{\delta\rho_\pm\left(\vec{r},t\right)}\right)\right]\,,\label{eq_DDFT}
\end{equation}
where $\rho_\pm\left(\vec{r},t\right)$ is the density of the ions with positive and negative charge respectively as a function of position $\vec{r}$ and time $t$; $\beta=1/k_BT$ is the inverse thermal energy and $D$ the diffusion coefficient of the ions which, for simplicity, is taken to be the same for both ion species $D_+=D_-\equiv D$. $\mathcal{F}$ denotes the free energy of the system which is a functional of the two densities.

For obtaining the time evolution of the densities from this equation knowledge of the free energy of the system is needed. It is given by the following expression, 
\begin{equation}
\mathcal{F}=\mathcal{F}^\textrm{id}+\mathcal{F}^\textrm{HS}+\mathcal{F}^\textrm{Coul}+\int d\vec{r}\ \rho_\pm\left(\vec{r},t\right)V_{\text{ext},\pm}\left(\vec{r},t\right)\,.\label{eq_Fcontri}
\end{equation} 
The ideal part $\mathcal{F}^\textrm{id}$ gives the free energy of an ideal gas. The remaining terms describe the interaction of the particles due to steric hard-sphere $\mathcal{F}^\textrm{HS}$ and charge effects $\mathcal{F}^\textrm{Coul}$ (Coulomb) as well as the effect of the external potential $V_{\text{ext},\pm}\left(\vec{r},t\right)=-q_\pm E(t) z$ describing the electric field
\begin{equation}
E(t)=E_0\cos(\omega t)
\end{equation} acting on the system. The electric field amplitude $E_0$ is connected to the voltage drop $\Delta U$ over the accessible system length by 
\begin{equation}
E_0=\Delta U/L\,.
\end{equation}

The ideal part $\mathcal{F}^\textrm{id}$ is known exactly, the hard-sphere part is described by \textit{fundamental measure theory} (FMT) \cite{Tarazona2000,Rosenfeld1989,RothReview2010,HaertelPhD,Oettel2010,Haertel2015JPCM} while the Coulomb interaction is taken into account with a mean-field approach \cite{HaertelPhD,Haertel2015EnEnvSci}, such that for the ions within a volume $V$ we obtain
\begin{align}
\mathcal{F}^\textrm{id}[\rho_+,\rho_-]&=k_BT\sum_{\pm}\int_{V}d\vec{r}\rho_{\pm}\left(\vec{r},t\right)\ln\left(\Lambda^3\rho_{\pm}\left(\vec{r},t\right)\right)\,,\nonumber\\
\mathcal{F}^\textrm{HS}[\rho_+,\rho_-]&=\mathcal{F}^\textrm{FMT}[\rho_++\rho_-]\,,\label{DDFT_Fterms}\\
\mathcal{F}^\textrm{Coul}[\rho_+,\rho_-]&=\frac{1}{8\pi\epsilon\epsilon_0}\int_{V} d\vec{r}\int_{V} d\vec{r}\,'\frac{\rho_c\left(\vec{r},t\right)\rho_c\left(\vec{r}\,',t\right)}{|\vec{r}-\vec{r}\,'|}\nonumber
\end{align}
with the charge density $\rho_c\left(\vec{r},t\right)=q\left(\rho_+\left(\vec{r},t\right)-\rho_-\left(\vec{r},t\right)\right)$. $\varepsilon_0$ denotes the vacuum and $\varepsilon$ the relative permittivity. $\Lambda$ is the de Broglie wavelength.

\subsubsection{Fundamental measure theory (FMT)}
We use the White Bear II version of fundamental measure theory in tensor form \cite{RothReview2010, HaertelPhD} to write the hard-sphere contribution to the free energy functional as 
\begin{equation}
\mathcal{F}^\mathrm{FMT}=\int_V d\vec{r}\Phi(\{n_\alpha\})\,.
\end{equation}
where $\Phi=\Phi_1+\Phi_2+\Phi_3$ is the free energy density with 
\begin{align}
\Phi_1&=-\frac{n_2}{4\pi R^2}\ln\left(1-n_3\right)\,,\nonumber\\
%
\Phi_2&=\frac{1}{4\pi R}\left(n_2^2-\vec{n}_2\cdot\vec{n}_2 \right)\frac{1+\frac{1}{3}\psi_2\left(n_3\right)}{1-n_3}\,,\\
\Phi_3&=\left(n_2^3-3 n_2\vec{n}_2\cdot\vec{n}_2+\frac{9}{2}\left(\vec{n}_2\te{n}_{m_2}\vec{n}_2-\mathrm{Tr}\left(\te{n}_{m_2}^3\right)\right)\right)\nonumber\\
&\qquad\times\frac{1-\frac{1}{3}n_3\psi_3\left(n_3\right)}{24\pi\left(1-n_3\right)^2}\nonumber\,,
\end{align}
using the functions
\begin{align}
\psi_2&=\frac{1}{n_3}\left(2n_3-n_3^2+2\left(1-n_3\right)\ln\left(1-n_3\right)\right)\,,\\
\psi_3&=\frac{1}{n_3^2}\left(2n_3-3n_3^2+2n_3^3+2\left(1-n_3\right)^2\ln\left(1-n_3\right)\right)\,.\nonumber
\end{align}
These expressions are solely dependent on a set of functions $n_\alpha$ referred to as weighted densities which are obtained from convolutions of the particle density $\rho$ with weight functions $\omega^{(\alpha)}$ such that
\begin{align} 
n_\alpha\left(\vec{r},t\right) 
=\int_{V}\rho\left(\vec{r}\,',t\right)\omega^{(\alpha)}\left(\vec{r}-\vec{r}\,'\right)d\vec{r}\,'\label{eq_weigthed_densities}
\end{align}
where
\begin{align}
\omega^{(2)}\left(\vec{r}\right)&=\delta\left(R-|\vec{r}|\right)\,,\nonumber\\
\omega^{(3)}\left(\vec{r}\right)&=\theta\left(R-|\vec{r}|\right)\,,\label{eq_weights}\\
%
\vec{\omega}^{(2)}\left(\vec{r}\right)&=\frac{\vec{r}}{|\vec{r}|}\delta\left(R-|\vec{r}|\right)\,,\nonumber\\
\overleftrightarrow{\omega}^{(m_2)}\left(\vec{r}\right)&= \left(\frac{\vec{r}\cdot{\vec{r}\,}^t}{|\vec{r}|^2}-\frac{\te{\mathbb{I}}}{3}\right)\delta\left(R-|\vec{r}|\right)\nonumber
\end{align}
are the weights in the case of spherical particles.
Here, $R=\sigma/2$ denotes the hard-sphere radius. $\delta$ is the Dirac delta function, $\theta$ the Heavyside step function, $\vec{r}\,^t$ is the transpose of the vector $\vec{r}$ and $\te{\mathbb{I}}$ is the unit matrix.

\subsection{System parameters and non-dimensionalization}
We will present parameterizations for the model on two different length scales.
For a micro-scale realization, we consider low-charged colloidal particles of charge $q=|q_\pm|=5\ e$, where $e$ is the elementary charge, and of diameter $\sigma=2.61$ $\mu$m which serves as the length scale. We further assume the ions to be partially solvent permeable with a hydrodynamic radius of $R_{\rm h}=\frac{\sigma}{20}$. The relative permittivity of the organic electrolyte is assumed as $\varepsilon=2.3$, corresponding for instance to the relative permittivity of a decalin-tetrachloroethylene mixture as discussed in Ref.\citenum{Chaudhuri2017}. The energy scale is set by the thermal energy at standard temperature ($T=298$ K), $k_BT=4.11\cdot 10^{-21}$ J, that can be used to define a thermal voltage \mbox{$U_{\rm T} = k_{\rm B} T/q = 5.14$ mV}. The viscosity of the mixture at room temperature is $\eta=1.29\cdot 10^{-3}$ Pa s, such that the diffusion constant of the macroions is $D_0=\frac{k_BT}{6\pi\eta R_{\rm h}}= 1.30\cdot 10^{-12}$ m$^2$ s$^{-1}$. We define a characteristic time scale \mbox{$\tau_0=\sigma^2/D_0= 5.25$ s} that corresponds to a characteristic angular frequency $\omega_0=\frac{2\pi}{\tau_0}=1.20$ s$^{-1}$.

On the other hand, a nano-scale system of relatively large monovalent ions at the same temperature with $q=e$, \mbox{$\sigma=2\ R_{\rm h}=3$ nm}, $\varepsilon=80$, \mbox{$U_{\rm T} = k_{\rm B} T/q = 25.7$ mV} leads to the same reduced system parameters. With the viscosity of water at room temperature, $\eta_\text{\rm water}=8.9\cdot 10^{-4}$ Pa s, the diffusion coefficient in this case is $D_0=1.63\cdot 10^{-10}$ m$^2$ s$^{-1}$ and the time and frequency scales are now given by \mbox{$\tau_0=5.5 \cdot 10^{-8}$ s} and $\omega_0=1.14\cdot 10^{8}$ s$^{-1}$.
The following calculation can thus be considered in either of these cases. A summary of the system parameters can be found in Table~\ref{table_syspara}.

The dimensionless amplitude of the external voltage is given by
\begin{equation}
U^*=\frac{\Delta U}{U_T} 
\end{equation}
As a baseline parameter for this amplitude, we use $U^* = 38.9$. \

In addition to the ion diffusion and self-diffusion time, $\tau_{\rm diff} = \frac{L^2}{D_0}$ and $\tau_{\rm 0} = \frac{\sigma^2}{D_0}$, we introduce the ion transit time, $\tau_{\rm tr}$ which corresponds to the time that an ion needs to cross the thickness of the device when the electric field is not screened. The latter is given by the ratio of the system length $L$ and the mean velocity of the particles in the unscreened case $\bar{v}=\frac{qE}{\gamma}\frac{2}{T}\int_{-T/4}^{T/4} dt \cos\left(\omega t\right)=\frac{2}{\pi}\frac{U^* D}{L}$ as $\tau_{\rm tr} = \frac{L}{\bar{v}}=\frac{\pi}{2} \frac{U_{\rm T}}{\Delta U} \frac{L^2}{D_0}$.
All time scales are normalized to the self-diffusion time $\tau_{\rm 0}$, such that the dimensionless ion transit time becomes 
\begin{equation}
\frac{\tau_{\rm tr}}{\tau_{\rm 0}} = \frac{\pi}{2} \frac{L^2}{\sigma^2U^*}
\end{equation}
and the dimensionless diffusion time of ions in the electrolyte is 
\begin{equation}
\frac{\tau_{\rm diff}}{\tau_{\rm 0}} = \frac{L^2}{\sigma^2}.
\end{equation}
The corresponding dimensionless angular frequencies are given by
\begin{equation}
\frac{\omega_{\rm tr}}{\omega_{\rm 0}} = \frac{2}{\pi} \frac{U^*\sigma^2}{L^2}\label{eq_omega_tr}
\end{equation}
and 
\begin{equation}
\frac{\omega_{\rm diff}}{\omega_{\rm 0}} = \frac{\sigma^2}{L^2}\,.
\end{equation}
The free parameters of the model, varied in simulations, are $\Delta U$ and $L$.

The dimensionless thickness $L/\sigma$ should be evaluated in relation to two common characteristic length scales of ionic systems: Bjerrum length $\lambda_B$ and Debye length $\lambda_D$, which are given in dimensionless form as  
\begin{align}
\frac{\lambda_{\rm B}}{\sigma}&=\frac{1}{4\pi\varepsilon}\ \frac{q^2}{\varepsilon_0k_BT\sigma}\,,\\
\frac{\lambda_D}{\sigma}&=\sqrt{\frac{\varepsilon\varepsilon_0 k_BT}{\sigma^2\sum_\pm n_\pm q_\pm^2}}=\sqrt{\frac{\varepsilon}{\frac{q^2}{\varepsilon_0 k_BT\sigma}}\frac{\pi}{6\phi}} = \frac{1}{2} \sqrt{\frac{\sigma}{\lambda_B} \frac{1}{6 \phi}}\nonumber\,,
\end{align}
where $n_\pm$ is the number density of cations and anions, respectively, and $\phi=\frac{\pi}{6}\sigma^3\left(n_++n_-\right)$ is the volume fraction occupied by ions, each defined on the accessible region of width $L$. We fix $\phi=0.375$ such that we obtain $\lambda_B/\sigma=0.23$ and $\lambda_D/\sigma = 0.69$ for the (macro-)ions. Thus both of the lengths, $\lambda_B$ and $\lambda_D$, are smaller than the particle diameter $\sigma$ and a strong effect of the hard-sphere interactions is expected.

\subsection{Quantities of interest}
At the start, the system is equilibrated without applied potential
to reach a steady state, at which point the transverse oscillating voltage $U(t)$ 
is introduced. The latter causes an electric field $E(t)$ and thus an ion flux in $z$-direction.
The resulting ion flux density $ \vec{j}_{p,\pm}\left(\vec{r},t\right)$ is a periodic function in time and can easily be derived from eq.~(\ref{eq_DDFT}) together with the continuity equation $\frac{\partial \rho_\pm\left(\vec{r},t\right)}{\partial t}=-\nabla \vec{j}_{p,\pm}\left(\vec{r},t\right)$ as
\begin{equation}
\vec{j}_{p,\pm}\left(\vec{r},t\right)=-\beta D\left[\rho_\pm\left(\vec{r},t\right)\nabla\left(\frac{\delta\mathcal{F}\left[\rho_+\left(\vec{r},t\right),\rho_-\left(\vec{r},t\right)\right]}{\delta\rho_\pm\left(\vec{r},t\right)}\right)\right]\,.\label{eq_j}
\end{equation}
The ionic current density in $z$-direction is then given as
\begin{equation}
j\left(\vec{r},t\right)=\sum_{\pm}q_\pm \vec{j}_{p,\pm}\left(\vec{r},t\right)\cdot\hat{\mathrm{e}}_z\,.\label{eq_current_dens}
\end{equation}
Analogously, we can define contributions to the ion flux density that stem from different energy terms, 
\begin{equation}
\vec{j}_{p,\pm}^k\left(\vec{r},t\right)=-\beta D\left[\rho_\pm\left(\vec{r},t\right)\nabla\left(\frac{\delta\mathcal{F}^k\left[\rho_+\left(\vec{r},t\right),\rho_-\left(\vec{r},t\right)\right]}{\delta\rho_\pm\left(\vec{r},t\right)}\right)\right]\,,\label{eq_jk}
\end{equation}
where $k\in\{\mathrm{id}, \mathrm{HS}, \mathrm{Coul}, \mathrm{ext}\}$
and the ionic current density contribution is then again given by eq.~(\ref{eq_current_dens}) replacing $\vec{j}_{p,\pm}$ with $\vec{j}_{p,\pm}^k$.
Note that considering the contributions due to the external potential, $\mathcal{F}^\textrm{ext}$, and the ideal term, $\mathcal{F}^\textrm{id}$, the equation is identical to the Nernst-Planck equation. Adding also the Coulomb interaction term, $\mathcal{F}^\textrm{Coul}$, we obtain the Poisson-Nernst-Planck (PNP) description. The additional hard-sphere interaction term, $\mathcal{F}^\textrm{HS}$, is a non-trivial extension to this model accounting for the finite ion size, an essential contribution due to the small system width.

The current response is non-dimensionalized relative to $j_0=e/\sigma^2\tau_0$ and can, due to its periodicity, be
decomposed into different harmonic contributions of amplitude $j_n$ with $n=1,2,3,\ldots$. The harmonics are obtained as
\begin{equation}
j_n\left(\vec{r}\right)=\frac{2}{T}\int_{-T/2}^{T/2} dt\ j\left(\vec{r},t\right)\mathrm{e}^{-in\omega t}
\end{equation}
with $T=2\pi/\omega$. Plots of the current will always show $j$ while we use $j_1$, the amplitude of the first harmonic of the current density, to define the complex impedance $Z$ given by 
%
%
\begin{equation}
\frac{Z}{Z_0}=U^*\ \frac{j_0}{j_1}\ \mathrm{e}^{i(\phi_U-\phi_{j_1})}\,.
\end{equation}
The scale of this quantity is set by $Z_0=k_BT\tau_0/ e^2$. Higher harmonic contributions to the current ($j_n$ with $n\ge2$) will be discussed in the Supplementary Information.

\begin{table*}[h!]
\renewcommand{\arraystretch}{1.2}
\caption{Summary of the system parameters. Typical values are given for a nanoscale system with $L=10$ nm, $\sigma=3$ nm, $q=e$, $D_0=1.63\cdot 10^{-10}$ m$^2$ s$^{-1}$, $k_BT=4.11\cdot 10^{-21}$ J, $\Delta U=1.0$ V and a microscale (colloidal) system $L=8.70$ $\mu$m, $\sigma=2.61$ $\mu$m, $q=5e$, $D_0= 1.30\cdot 10^{-12}$ m$^2$ s$^{-1}$, $k_BT=4.11\cdot 10^{-21}$ J, $\Delta U=0.2$ V, both leading to the same normalized system parameters.}\label{table_syspara}
\centering
\begin{tabular}{ p{3.5cm}   c  p{4.0cm}  p{2.8cm}  p{2.5cm} p{2cm}}
dimensional property & symbol & definition & normalized value& typical value \newline nanoscale & typical value \newline microscale\\ \hline 
ion diameter &$\sigma$ & & 1 & $3$ nm & $2.61$ $\mu$m\\
accessible system length &$L$ & & $L/\sigma$ & $10$ nm & $8.70$ $\mu$m\\
Bjerrum length &$\lambda_B$& $q^2/(4\pi\varepsilon\varepsilon_0k_BT)$& $q^2/(4\pi\varepsilon\varepsilon_0k_BT\sigma)$& $0.70$ nm & $0.61$ $\mu$m\\
Debye length &$\lambda_D$& $\frac{\sigma}{2} \sqrt{\frac{\sigma}{\lambda_B} \frac{1}{6 \phi}}$&  $\frac{1}{2} \sqrt{\frac{\sigma}{\lambda_B} \frac{1}{6 \phi}}$ &$2.07$ nm &$1.80$ $\mu$m\\ \hline
self-diffusion time & $\tau_0$ &  $\sigma^2/D_0$ & 1 & $5.50 \cdot 10^{-8}$ s & $5.25$ s\\
driving period &$T$ &  & $T D_0/\sigma^2$& &\\
transit time &$\tau_{\rm tr}$ & $\pi L^2k_BT/(2qUD_0)$ &  $\pi L^2 /(2\sigma^2U^*)$& $2.47 \cdot 10^{-9}$ s & $2.35$ s\\
diffusion time &$\tau_{\rm diff}$ &  $L^2/D_0$ & $L^2/\sigma^2$& $6.12 \cdot 10^{-7}$ s & $58.3$ s\\ \hline 
self-diffusion frequency & $\omega_0$ & $2\pi D_0/\sigma^2$& $2\pi$  & $1.14\cdot 10^{8}$ s$^{-1}$ & $1.20$ s$^{-1}$\\
driving frequency &$\omega$ &  & $\omega\sigma^2/2\pi D_0$& &\\
transit frequency &$\omega_{\rm tr}$ & $4qUD_0/( L^2k_BT)$& $4\sigma^2 U^*/L^2$ & $2.54 \cdot 10^{9}$ s$^{-1}$ & $2.67 $ s$^{-1}$\\
diffusion frequency &$\omega_{\rm diff}$ & $2\pi D_0/L^2$ & $2\pi\sigma^2/L^2$& $1.03\cdot 10^{8}$ s$^{-1}$ & $0.11$ s$^{-1}$\\ 
\hline 
current scale &$j_0$ & $q/(\sigma^2\tau_0)$ & 1 & $3.23\cdot10^5$ A/m$^2$ & $2.24\cdot10^{-8}$ A/m$^2$\\ 
current harmonics &$j_n$ & $\frac{2}{T}\int_{-T/2}^{T/2} dt\ j\left(\vec{r},t\right)\mathrm{e}^{-in\omega t}$ & $j_n/j_0$ & &\\ \hline
thermal voltage &$U_T$ & $k_BT/q$ & 1 & $25.7$ mV & $5.14$ mV\\ 
external voltage &$\Delta U$ & $\Delta U= E_0L$ & $\Delta U/U_T=U^*$ & $ 1$ V & $ 0.2$ V\\ \hline
impedance scale &$Z_0$ & $k_BT\tau_0/ q^2$ & 1 & $8.8\cdot 10^9$ $\Omega$ & $3.36\cdot 10^{16}$ $\Omega$\\
impedance &$Z$& $U/(j_1\sigma^2)$& $U^*j_0/j_1$& &\\
\hline
relative permittivity  &$\epsilon$ &  &  &$80$ & $2.3$\\ 
particle charge &$q$ & &$q/e$ & $1e$ & $5e$
\end{tabular}
\end{table*}

\section{Results and discussion}

\subsection{Density profiles and current response}
\begin{figure}[h!]
\centering
\includegraphics[width=0.8\columnwidth]{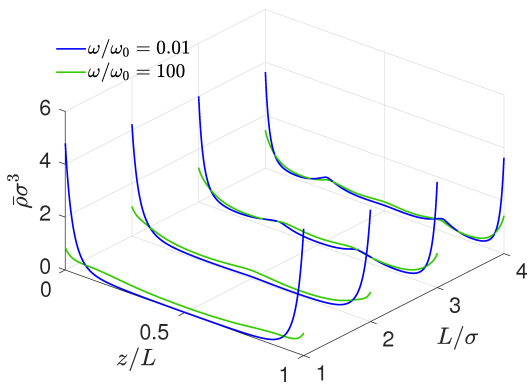}
\caption{Mean density distribution $\bar{\rho}\equiv\bar{\rho}_+=\bar{\rho}_-$ averaged over one period for different widths, $L/\sigma=1$, $L/\sigma=2$, $L/\sigma=3$, $L/\sigma=4$. Layering at the walls and at multiples of $\sigma$ is observed. The effect is stronger for smaller frequencies.}\label{fig_density_profiles}
\end{figure}

\begin{figure}[h!]
\centering
\includegraphics[width=0.7\columnwidth]{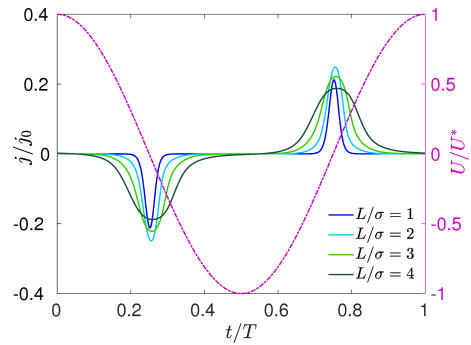}
\caption{Time dependence of current density at the center position ($z=L/2$) at small frequency ($\omega/\omega_0=0.01$). The resulting current shows strong anharmonicity and a phase shift with respect to the driving voltage $U(t)$ (curve plotted as purple dash-dot line against right $y$ axis for reference).}\label{fig_current_at_smallfreq}
\end{figure}

First, we investigate the effect of the system width on the density distribution and current induced in the system.
In fig.~\ref{fig_density_profiles} the effect of the hard-sphere interaction is reflected in the increased values of the particle densities at the walls and at multiples of $\sigma$. Counterion accumulation at the walls is enhanced with the application of a finite voltage difference and for larger system widths, when ions are attracted to walls of opposite charge. The effect is preserved but weakens as the frequency increases. Thus, lower frequencies correspond to higher amounts of localized ions at the walls. Further, while at high frequencies the current is harmonic and in phase with the driving voltage, the low frequency current shows anharmonicity and a phase-shift, see fig.~\ref{fig_current_at_smallfreq}.
For stronger electric field (smaller system lengths) the current becomes increasingly peaked and the system response is non-linear.



\subsection{Impedance response}
To examine the effect of the hard-sphere character of the ions, eq.~(\ref{eq_DDFT}) was solved, for reference, with and without hard-sphere (HS) and Coulomb (Coul) interaction terms and the impedance corresponding to the ion flux at the capacitor midplane was calculated, see fig.~\ref{fig_impedance_vs_w}. Hard-sphere interactions lead to a large increment of this impedance at medium and high frequencies, whereas the effect of Coulomb interactions in determining the impedance is much less pronounced.

The hard-sphere contribution is responsible for a maximum in the impedance at $\omega > \omega_{\rm tr}$. 
This feature vanishes, when the hard-sphere contribution is switched off. 
We conjecture that the maximum is thus related to the additional structure in the density distribution induced by hard-sphere interactions.
 

Another peculiar feature in fig.~\ref{fig_impedance_vs_w} is the impedance minimum seen at $\omega \approx \omega_0$. This feature is independent of the hard-sphere character of ions and also independent of their Coulomb interaction. It represents a \textit{resonance phenomenon} that should be common to all systems of confined ions exposed to an oscillating external potential. However, observation of this phenomenon depends critically on system parameters. It is a peculiar signature of the pronounced wall effects, which prevail in strongly confined systems upon application of an AC voltage with large amplitude. Under normal conditions in planar capacitive devices
the resonance should be quenched by thermal diffusion \cite{Beunis2008, HaertelReview2017, Bazant2004, Bazant2010, Feicht2016}. Diffusion causes a melting or dephasing of the highly coherent ion motion induced by wall effects. Since the occurrence of the impedance resonance is not affected by hard-sphere or Coulombic interaction terms, the phenomenon can be illustrated and explained using a highly simplified model, which will be presented next.

\subsection{Resonance effect} 
\subsubsection{Single-ion capacitor model}

\begin{figure}[htb]
\includegraphics[width=0.45\columnwidth]{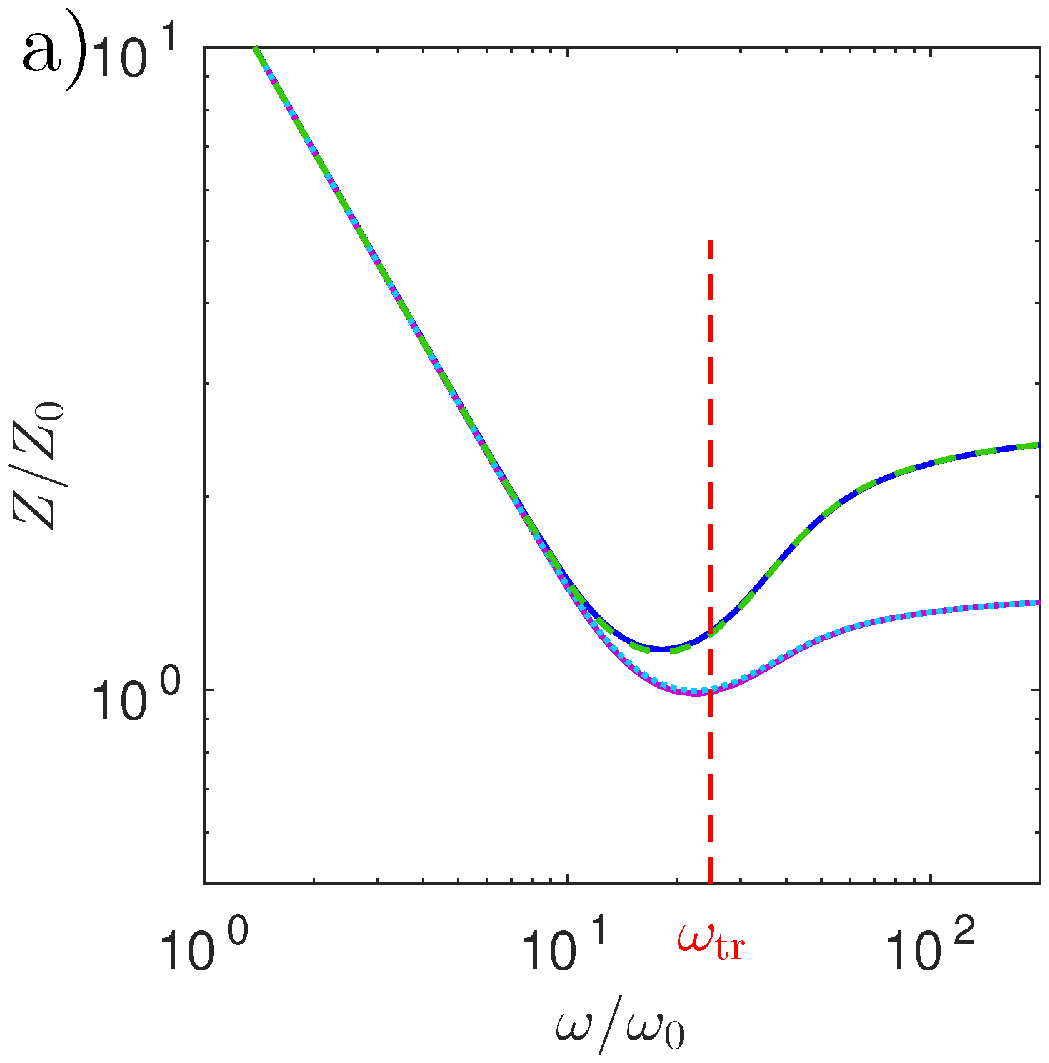}\hfill
\includegraphics[width=0.45\columnwidth]{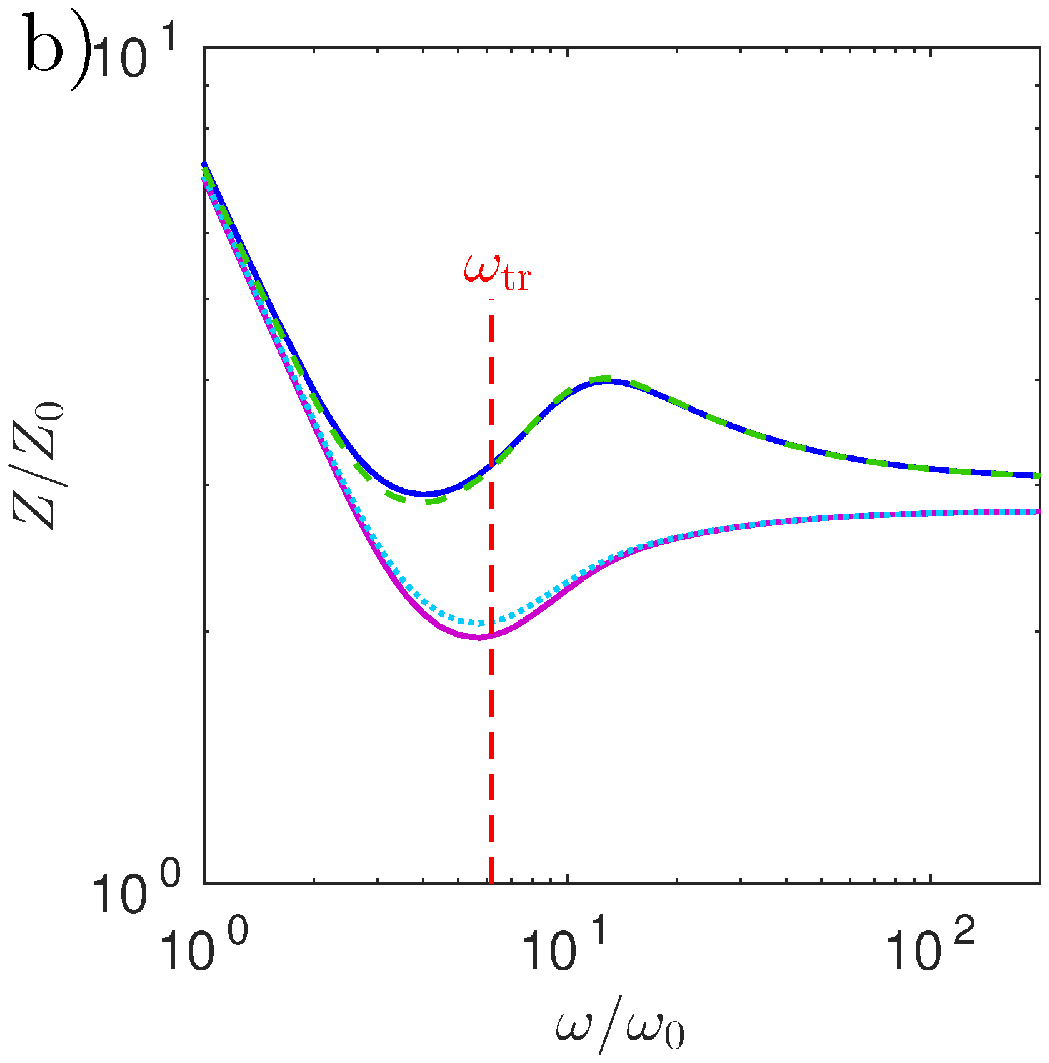}
\includegraphics[width=0.45\columnwidth]{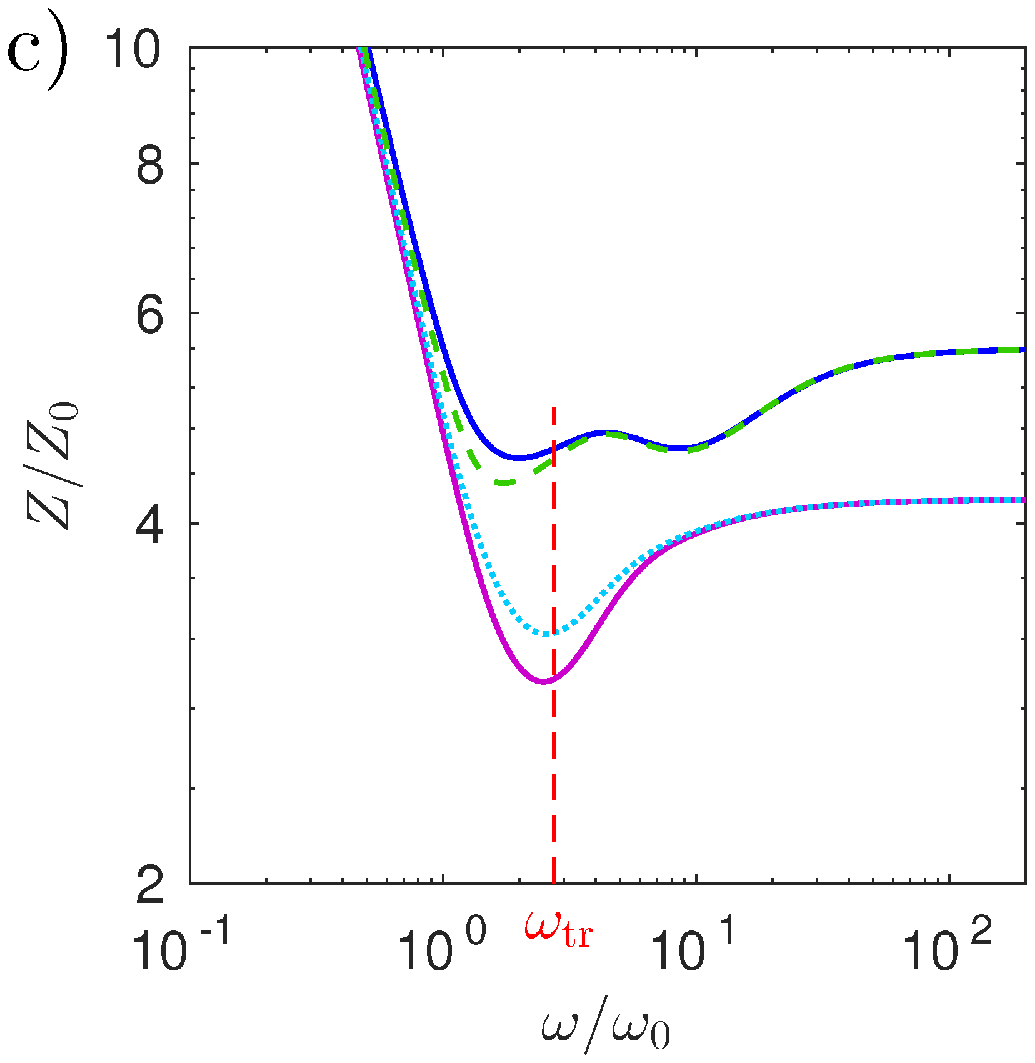}\hfill
\includegraphics[width=0.45\columnwidth]{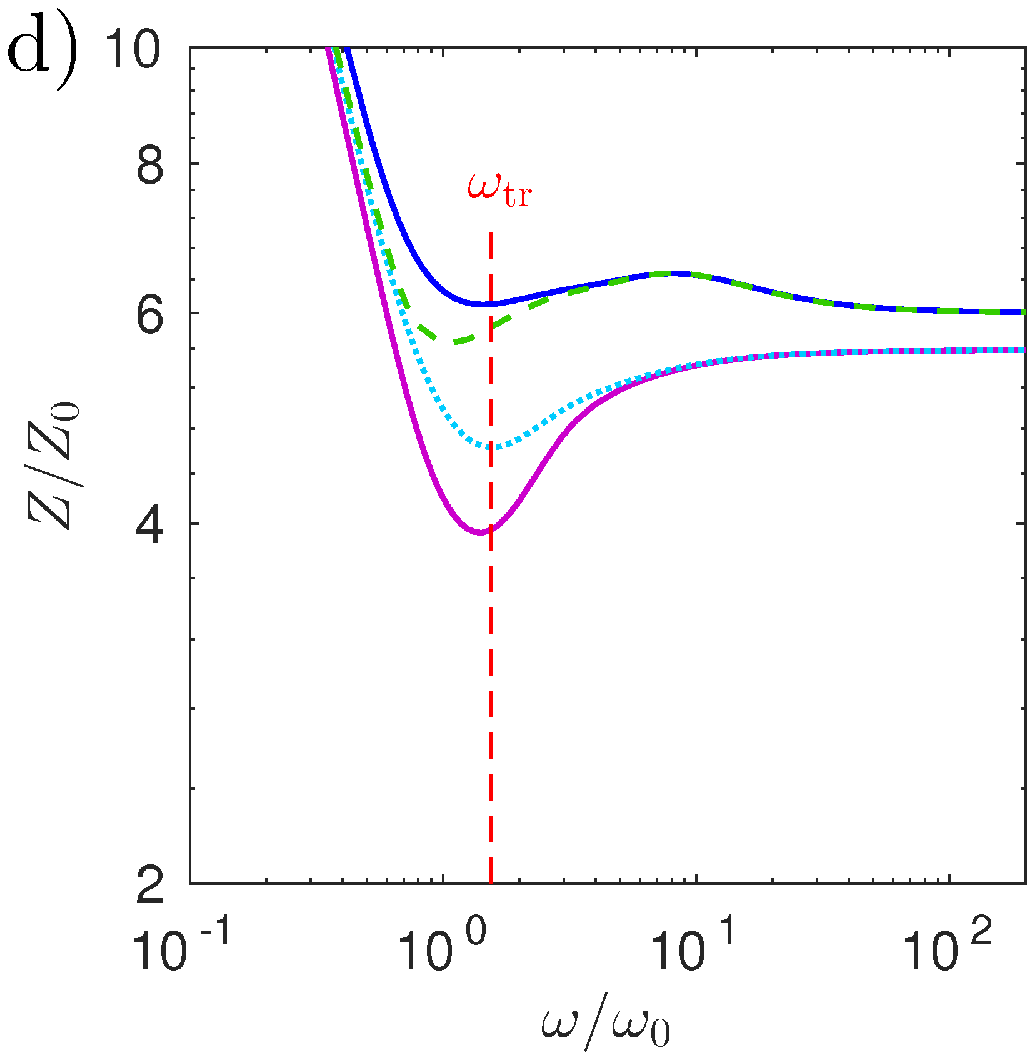}
\flushleft
\includegraphics[width=0.75\columnwidth]{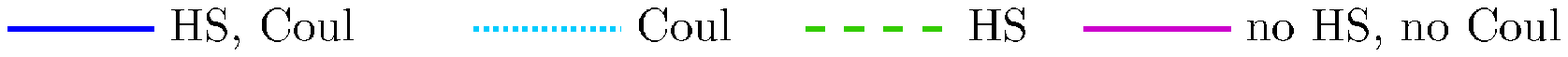}
\caption{Impedance with respect to angular frequency of the driving voltage $\omega$ for systems of lengths (a) $L/\sigma=1$, (b) $L/\sigma=2$, (c) $L/\sigma=3$, (d) $L/\sigma=4$. The large frequency limit impedance values are reached from below for system lengths that are odd multiples of $\sigma$ and from above for even ones. The analytic result for the position of the minimum in the impedance $\omega_{\rm tr}$ according to eq.~(\ref{eq_omega_tr}) is also indicated.}\label{fig_impedance_vs_w}
\end{figure}

The origin of the minimum in fig.~\ref{fig_impedance_vs_w} must be universal and can thus be understood for the simple case of a non-interacting gas of ions with charge $q$ confined between two charged plates. For simplicity we neglect thermal motion.  In the non-interacting case every ion can be considered individually. The equation of motion for each ion is equivalent to the case of a capacitor configuration with just one ion between the plates and is given by
\begin{equation}
\dot{z}(t)=\frac{qE_0}{\gamma}\cos(\omega t)\label{eqm}
\end{equation}
with the friction coefficient $\gamma$ and the angular frequency $\omega$.
The ion position is then
\begin{equation}
z(t)=z_0+\frac{qE_0}{\omega\gamma}\sin(\omega t)
\end{equation}
with arbitrary starting position $z_0$. During one oscillation period, an ion will transfer through a total transverse distance $d_{\rm tr} =2qE_0/\omega\gamma$ between the plates. 

In terms of the distance $d_{\rm tr}$ and frequency $\omega$, we can distinguish three regimes. In the regime of small amplitude, $d_{\rm tr} < L/2$, and high frequency, $\omega > 4qE_0/\gamma L$
, ions that cross the midplane and contribute to $j$ and $Z$ at this plane, perform full harmonic oscillations with zero phase shift to the applied AC voltage. The resulting ionic current density is easily obtained as
\begin{equation}
j(z=L/2,t)=q\dot{z}(t)\rho_0=\frac{q^2E_0\rho_0}{\gamma}\cos\left(\omega t\right)\label{eq_j_highf}
\end{equation}
in the case of negligible thermal motion.

In the regime of large amplitude, ${d}_{\rm tr} > L$, and small frequency, $\omega < 2qE_0/\gamma L$
, all ions will accumulate at either one of the surface planes during a half-period. Therefore, under ideal conditions, as considered with this simple model, cations and anions will perform a highly coherent motion and cross the midplane as two condensed and oppositely directed layers, see also fig.~\ref{fig_current_at_smallfreq}. In this regime, all ions will contribute to current and impedance responses determined at this plane. The ionic current density, averaged over a half period, will thus be proportional to the frequency of the applied field; it will be highly anharmonic and exhibit a monotonically decreasing phase shift with decreasing $\omega$, approaching $-\frac{\pi}{2}$ in the zero frequency limit. The solution of eq.~(\ref{eqm}) in this and the following case are given in the Supplementary Information.


In the intermediate regime with $L/2 \le d_{\rm tr} \le L$ and 
$\frac{4q E_0}{\gamma L} \ge \omega \ge  \frac{2 qE_0}{\gamma L}$, all ions in the system contribute to current density and impedance at the midplane; however, a fraction $d_{\rm tr}/L$ of these ions form a condensed layer at the walls, whereas the remaining fraction of ions remains distributed uniformly in between and follows the applied field harmonically and with zero phase shift. The transition that gives rise to the impedance resonance occurs at $\omega_{\rm tr} = \frac{4q E_0}{\gamma L}$: slightly above this frequency, only 50\% of ions (corresponding to $d_{\rm tr}/L$) contribute the ion flux at the midplane and as the frequency increases, this fraction diminishes with the decrease of $d_{\rm tr}/L$. In the frequency range at and below $\omega_{\rm tr}$, 100\% of ions contribute to the midplane current, as a consequence of the ion condensation at the walls. Thus the resonance seen in fig.\ref{fig_impedance_vs_w} has a simple geometric interpretation. 

A necessary condition for observing this resonance at finite temperature is that diffusional dephasing of the coherent ion motion will take place on a time scale that is much larger than the ion transit time, i.e., $\tau_{\rm diff} \gg \tau_{\rm tr}$ or $\omega_{\rm diff} \ll \omega_{\rm tr}$. The critical parameter that decides about this condition is $U^*$, which should be much larger than 1 for the ion condensation effect to be discernible.

Further, the preceding small amplitude or high frequency case, eq.~(\ref{eq_j_highf}), can be adopted to determine the high frequency limit behavior of non-uniform distributions of interacting particles by interpreting $\rho$ as a local density, which we understand as the mean density over the length that particles oscillate, $d_{\rm tr}$.
Then, if the local density has a maximum at the plane of interest, $j$ grows the smaller $d_{\rm tr}$, i.e.~the larger $\omega$, and the impedance $Z\propto1/j$ declines towards its high frequency value. For a minimum in the density the opposite is true and we approach the constant high-frequency impedance from below. This effect is also visible in fig.~\ref{fig_impedance_vs_w} for interacting particles (HS, Coul). For system lengths that are even multiples of $\sigma$, there is a maximum in the density at the center position when hard-sphere interactions are included and the high frequency limit of the impedance is approached from above. For odd multiples of $\sigma$, the center position is at a density minimum and the high frequency limit of the impedance is approached from below. 
Similarly, density inhomogeneities in the vicinity of the considered plane also appear in the impedance response at corresponding frequencies leading to additional extrema at medium frequencies.

\subsubsection{Rescaled resonance}
\begin{figure}[htb]
\centering
\includegraphics[width=0.65\columnwidth]{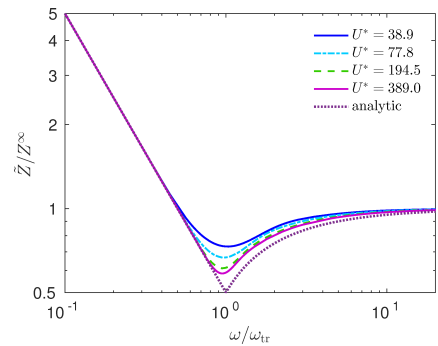}
\caption{Impedance $\tilde{Z}$ for different values of the external voltage $U^*$ and for the approximative analytic result with respect to angular frequency $\omega$. The results are rescaled to the transition frequency $\omega_{\rm tr}$ and the high frequency impedance limit $Z^\infty$.
The observed resonance at $\omega/\omega_{\rm tr}=1$ becomes more pronounced for higher $U^*$.}\label{fig_imedance_rescaled}
\end{figure}

We use the one-particle model to further investigate the emerging resonance. Defining the current averaged over a period of the driving signal $\bar{j}=\frac{1}{T}\int_{0}^{T}dt j(t)$, we determine the corresponding time-averaged impedance $\tilde{Z}=\Delta U/\bar{j}\sigma^2$. 
For high frequencies, $j$ is given by eq.~(\ref{eq_j_highf}) in the athermal case. At \textit{finite} temperature we find that even though the condensed particles at the walls do not contribute to the current directly, diffusion from the condensed part into the gap center will lead to a density higher than the equilibrium density $\rho_0$ there. This effect is particularly dominant for medium frequencies where the fraction of condensed particles, $d_\text{tr}/L$, is expected to be high. As a next order improvement, we correct for this effect by neglecting the peaked structure altogether. The numerical results for the density show that this is a valid approximation as the ion density peaks account for only about 10\% of all ions. Considering only the uniform distribution part, the leftmost ions reach the extremal position $d_{\rm tr}$ while the rightmost are pushed against the wall at position $L$. The effective width available to the ions is thus reduced to $L-d_{\rm tr}$. The density, assuming again a constant distribution but now over the reduced region, is given by 
\begin{equation}
\rho=\rho_0L/(L-d_{\rm tr})\,.\label{eq_rho_redarea}
\end{equation}
From eq.~(\ref{eq_j_highf}) and (\ref{eq_rho_redarea}) we thus obtain 
\begin{equation}
\bar{j}=\frac{2}{\pi}\frac{q^2E_0}{\gamma}\frac{\rho_0L}{L-d_{\rm tr}}\label{eq_jbar_rhomod}
\end{equation}
and the high frequency limit, $\omega\rightarrow\infty$, as $j^\infty=\frac{2}{\pi}\frac{q^2E_0\rho_0}{\gamma}$.
Normalizing the frequencies to the transit frequency $\omega_{\rm tr}$ and 
the impedance to the high frequency limit $Z^\infty=U^*j_0/j^\infty$, we find that the result is solely depending on the value of $U^*$, see fig.~\ref{fig_imedance_rescaled}. An approximate analytic result in the limit of negligible thermal motion compared to the driving force is given by eq.~(\ref{eq_jbar_rhomod}) for $\omega>\omega_{\rm tr}$ and $\bar{j}=q\rho_0L\omega/\pi$ for $\omega\le\omega_{\rm tr}$ when all ions are passing the midplane.
The resonance becomes more pronounced for higher values of $U^*$. However, it cannot exceed a factor of $2$ between the resonance value and the high frequency limit of the impedance.
High values of $U^*$ may appear unphysical but could be realized by using multivalent ions rather than higher voltages.

\subsection{Effect of condensed layer}

\subsubsection{Impedance}
So far we have only considered the impedance corresponding to the current at the system center at $z=L/2$. However, it is intuitive to expect the local impedance response, i.e. the impedance associated with the time-dependent current at a fixed point between the capacitor plates, to be highly dependent on the position. The current response in the ion layer at the wall should significantly differ from the current in the system center. As a next step we therefore consider the dependence of the local impedance on the position between the two capacitor plates.
\begin{figure}[htb]
\centering
\includegraphics[width=0.45\columnwidth]{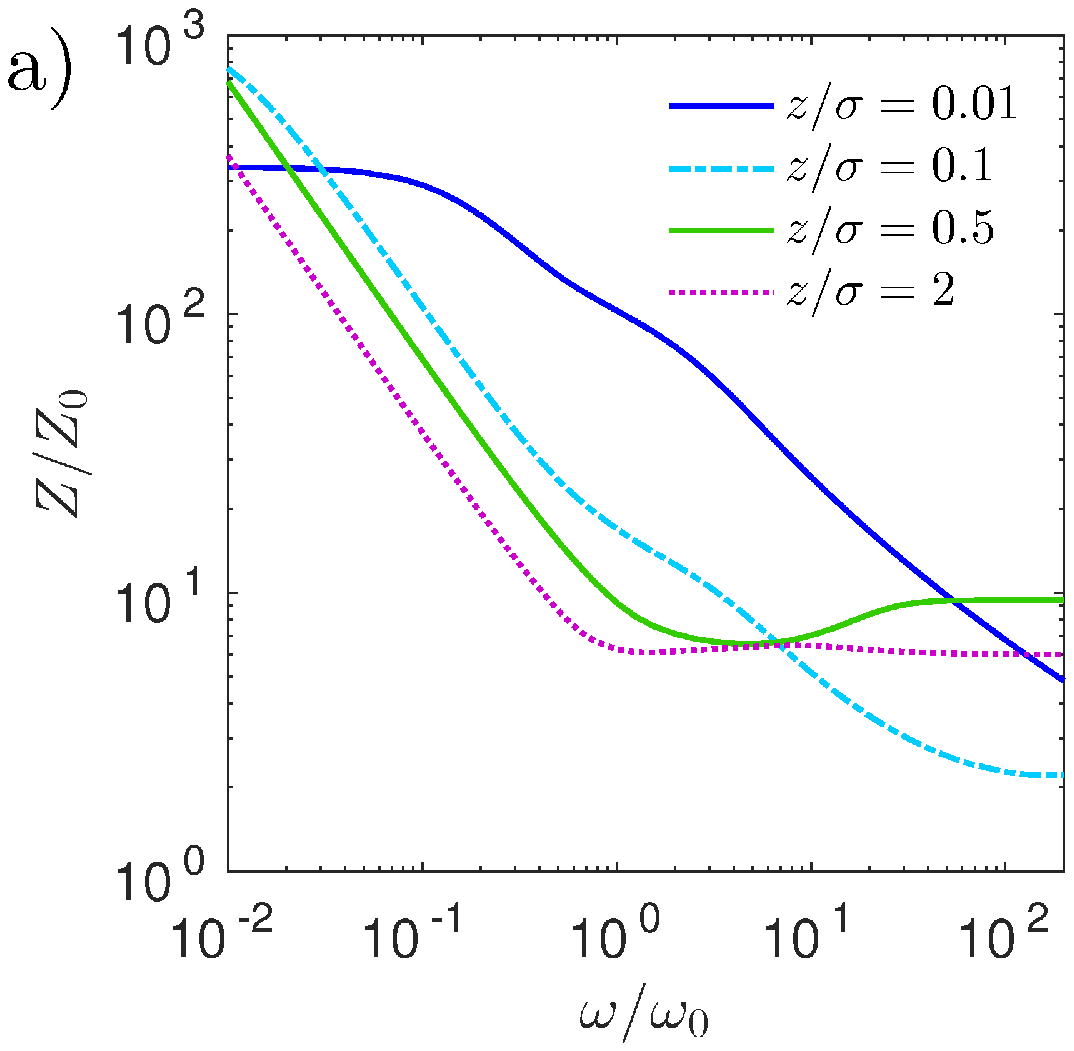}\hfill
\includegraphics[width=0.45\columnwidth]{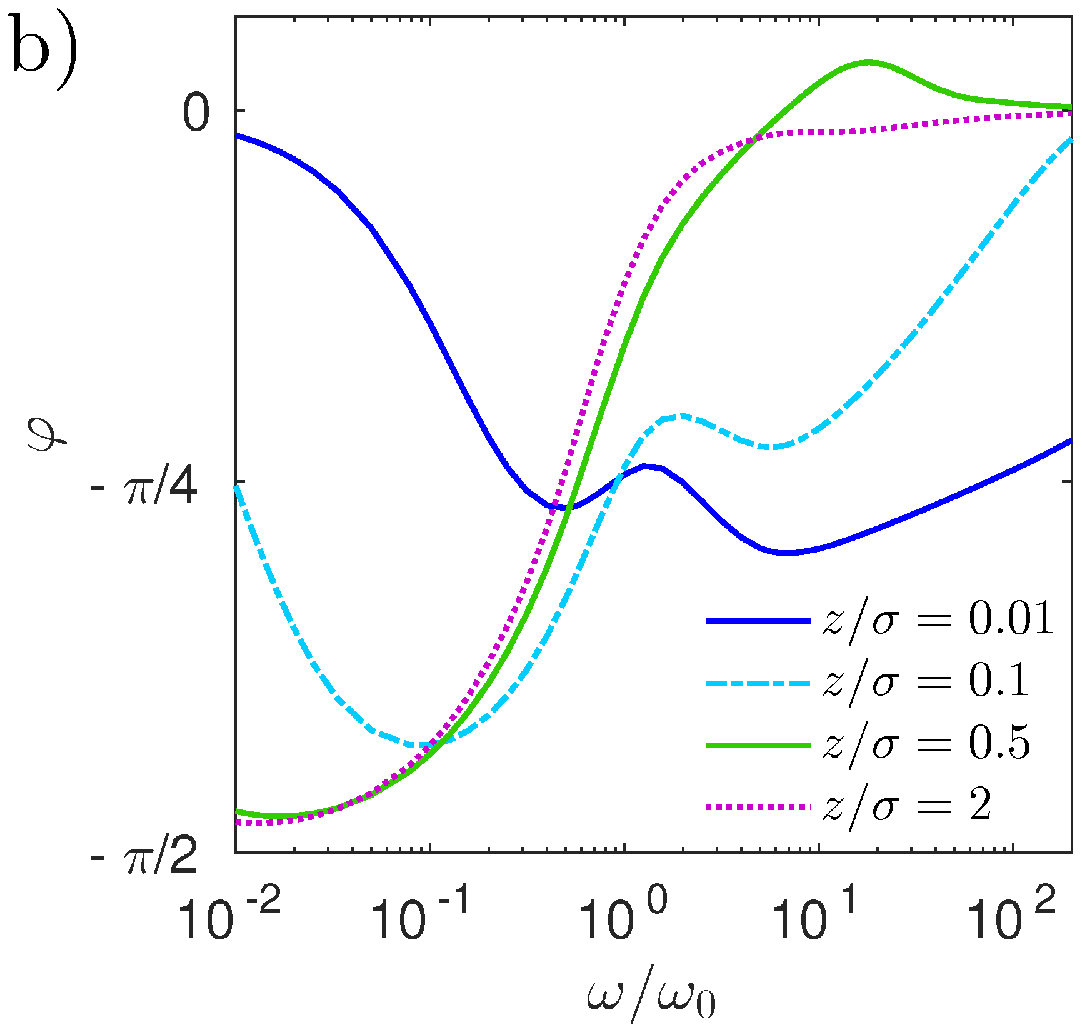}
\caption{Impedance $Z$ (a) and phase $\varphi$ (b) as functions of frequency for different positions $z$ between the plates. The frequency dependent impedance changes qualitatively when approaching the wall. The parameters are $L/\sigma=4$, $U^*=38.9$.}\label{fig_Impedance_dep_pos}
\end{figure}
In fig.~\ref{fig_Impedance_dep_pos} the case $z/\sigma=2$ corresponds to the center of the system. The system response here is as expected with a $1/\omega$ decay to the high frequency constant value in the impedance amplitude and the phase changing from $-\pi/2$ to $0$. Approaching the wall, the position of $z/\sigma=0.5$ corresponds to a density minimum and the large frequency limit of the impedance is approached from below. The rise in $Z$ towards the high frequency limit 
is accompanied by a maximum in the phase with the system response even becoming inductive ($\varphi>0$) for a limited frequency range. The phase behavior is caused by the ion condensation at the wall. These ions lead to a temporal shift of the maximum in the current response towards the time at which they pass the plane. The resulting phase shift increases with $\omega$ and may even become positive. Upon further increasing $\omega$, ions from the wall no longer reach the plane and the effect abates.

Decreasing the distance to the wall, we find that the low frequency phase value approaches zero as we enter the region of condensed ions. 
If we now consider again the time the ions condensed at the wall take to reach the plane at which we determine the impedance, we find that some of these ions are already present at the considered plane 
reducing the phase shift towards zero. The effect becomes stronger the more ions are present at the plane, so the closer we are to the wall. 
At large frequencies the oscillation amplitude is very small and the current is constituted by the ions freely oscillating in the field, thus also in this limit the phase grows towards zero.

\subsubsection{Current components}

\begin{figure}[htb]
\centering
\includegraphics[width=0.49\columnwidth]{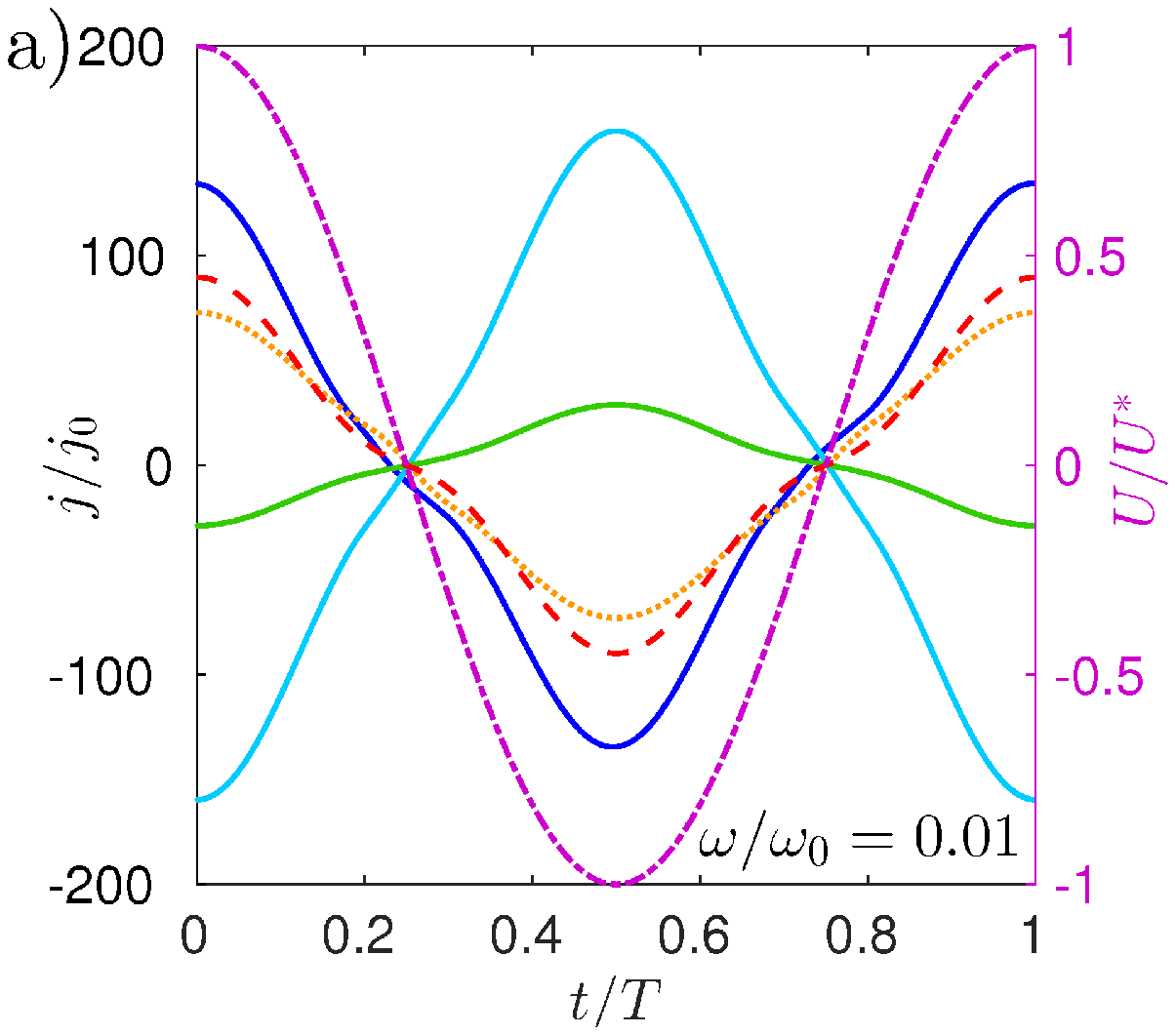}
\includegraphics[width=0.49\columnwidth]{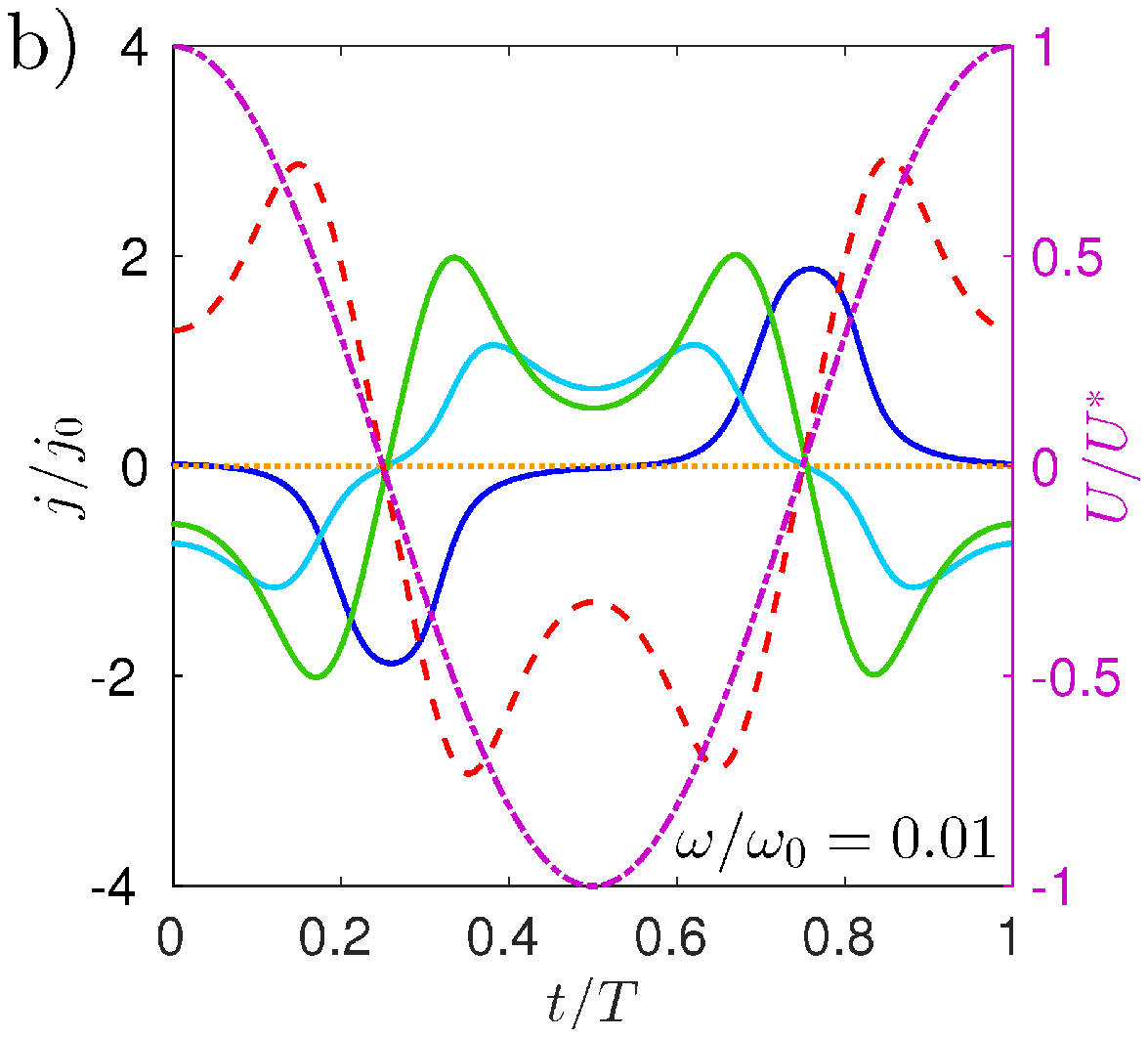}
\includegraphics[width=0.49\columnwidth]{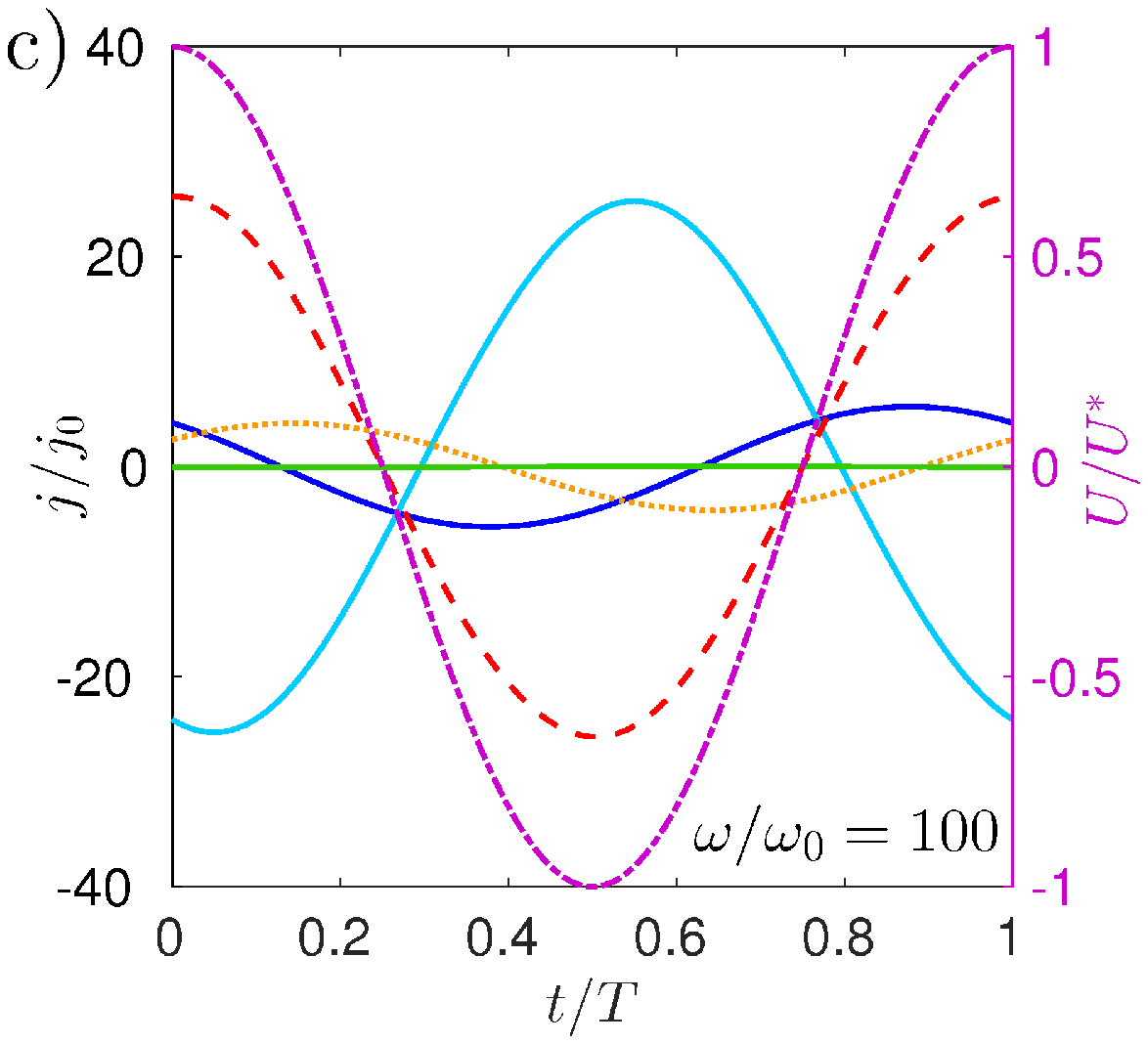}
\includegraphics[width=0.49\columnwidth]{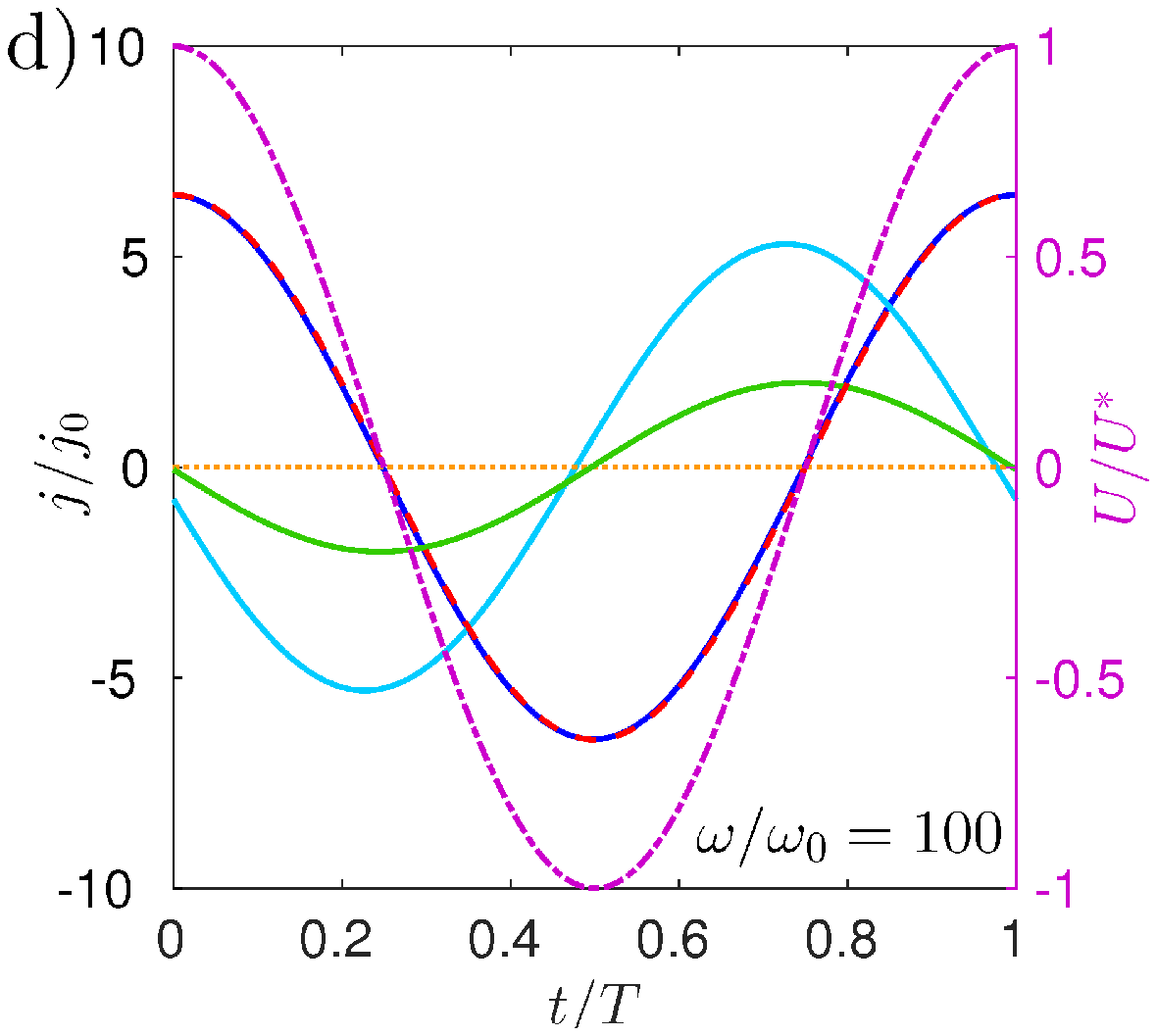}
\flushleft\includegraphics[width=0.75\columnwidth]{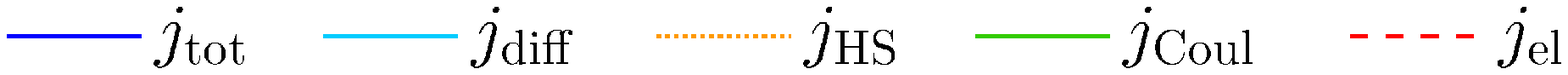}
\caption{Contributions of the different flow terms to the system response for a system of length $L/\sigma=4$  near the wall(a, c) ($z/\sigma=0.01$) and at the center (b, d) ($z/\sigma=2$). The contributions to the currents show considerable anharmonicity at small frequencies (upper row) which disappears at high frequencies (lower row). The effect of particle interaction grows for high densities close to the wall, and small frequencies. For better visibility the current components have been scaled in (a) $j_\text{tot}\times 1000$, $j_\text{Coul}\times 10$, (b) $j_\text{tot}\times 10$, (d) $j_\text{diff}\times 100$, $j_\text{HS}\times 100$, $j_\text{Coul}\times 100$. For reference the voltage  $U(t)$ is plotted against the right $y$ axis.}\label{fig_current_components}
\end{figure}
To better understand the effect of ion interaction on the current, we separate the flow into contributions corresponding to the different energy terms in eq.~(\ref{eq_Fcontri}). In fig.~\ref{fig_current_components}, the flows caused by the external electric field, $j_\mathrm{el}$, by the hard-sphere, $j_\mathrm{HS}$, and Coulomb interaction, $j_\mathrm{Coul}$, and the one due to diffusion of the ions, $j_\mathrm{diff}$, are shown together with the total flow, $j_\text{tot}$, as a function of time for two different frequencies. 

At the center (right column) the largest contribution to the total flow is always given by the flow due to the external electric field. Owing to the moderate density at this position the hard-sphere interaction term is negligible. At high frequencies the ions are freely following the electric field with a small amplitude oscillation around their position. The oscillation is in phase with the external electric field and the effect of the interaction terms is minute. For lower frequencies, however, the amplitude of the oscillatory motion becomes larger and the ions feel their neighbors via their Coulomb potential. Together with the diffusive flow the Coulomb effect counteracts the particle oscillation induced by the electric field resulting in a phase-shifted total flow $j_\text{tot}$.

Close to the wall, and at low frequencies, there are three main components that enter the total flow. These are given by the hard-sphere interaction, the diffusion and the external field term. 
The external electric field causes high densities near the walls for which the hard-sphere interaction becomes important. Additionally, the high density gradient close to the walls causes a strong diffusive flow. The electric field and hard-sphere terms lead to charge flows in the same direction, pushing the ions against the wall. This flow is countered by the diffusive flow which works against the formation of the accompanying density gradient. The different contributions show large anharmonicity and the resulting current is tiny. At high frequencies the current contributions are harmonic and the hard-sphere flow term is reduced due to the smaller oscillation amplitude.

\section{Conclusions}
We have explored the effect of an alternating voltage on a symmetric binary electrolyte confined between plan-parallel
capacitor plates by dynamic density functional theory \cite{Evans2016,Archer2004,Espanol2009,Marconi1999}.
This nonlinear theory provides a unifying framework to include steric interactions
between ions and their drift dynamics induced by the AC electric field. It thus significantly generalizes previous approaches like the
traditional Poisson-Nernst-Planck theory. Besides the dynamical layering of the driven ions near the confining walls, we predict
a resonance effect of the impedance which can be traced back to a simple single-ion effect but is modified by Coulomb
and steric interactions. This effect becomes relevant in nano-confinement
and allows to tune the electric response by confinement spacing, temperature and external voltage applied.

Future work should concentrate on the molecular details of the solvent as well as specific substrate properties like surface charges and roughness. In particular, reorientation and polarization effects of the liquid molecules and the impact of these effects on the capacitive ion response discussed here should be considered.
Density functional theory can in fact be generalized towards solvent-ion mixtures \cite{Oleksy2009,Medasani2014} and more general external potentials. Modeling of electrode details can in principle be treated as well.

%
%

\section*{Acknowledgments}
We thank Alexei A. Kornyshev and Andreas H\"artel for helpful discussions. We gratefully acknowledge financial support by the DFG grant LO 418/19-1.

\section*{Supplementary Information}


The following files are available free of charge.
\begin{itemize}
  \item Supplemental1.pdf: Current form for medium and low frequencies; Higher moments of the current
\end{itemize}


\bibliography{references_arxiv}

\begin{thebibliography}{88}%
\makeatletter
\providecommand \@ifxundefined [1]{%
 \@ifx{#1\undefined}
}%
\providecommand \@ifnum [1]{%
 \ifnum #1\expandafter \@firstoftwo
 \else \expandafter \@secondoftwo
 \fi
}%
\providecommand \@ifx [1]{%
 \ifx #1\expandafter \@firstoftwo
 \else \expandafter \@secondoftwo
 \fi
}%
\providecommand \natexlab [1]{#1}%
\providecommand \enquote  [1]{``#1''}%
\providecommand \bibnamefont  [1]{#1}%
\providecommand \bibfnamefont [1]{#1}%
\providecommand \citenamefont [1]{#1}%
\providecommand \href@noop [0]{\@secondoftwo}%
\providecommand \href [0]{\begingroup \@sanitize@url \@href}%
\providecommand \@href[1]{\@@startlink{#1}\@@href}%
\providecommand \@@href[1]{\endgroup#1\@@endlink}%
\providecommand \@sanitize@url [0]{\catcode `\\12\catcode `\$12\catcode
  `\&12\catcode `\#12\catcode `\^12\catcode `\_12\catcode `\%12\relax}%
\providecommand \@@startlink[1]{}%
\providecommand \@@endlink[0]{}%
\providecommand \url  [0]{\begingroup\@sanitize@url \@url }%
\providecommand \@url [1]{\endgroup\@href {#1}{\urlprefix }}%
\providecommand \urlprefix  [0]{URL }%
\providecommand \Eprint [0]{\href }%
\providecommand \doibase [0]{http://dx.doi.org/}%
\providecommand \selectlanguage [0]{\@gobble}%
\providecommand \bibinfo  [0]{\@secondoftwo}%
\providecommand \bibfield  [0]{\@secondoftwo}%
\providecommand \translation [1]{[#1]}%
\providecommand \BibitemOpen [0]{}%
\providecommand \bibitemStop [0]{}%
\providecommand \bibitemNoStop [0]{.\EOS\space}%
\providecommand \EOS [0]{\spacefactor3000\relax}%
\providecommand \BibitemShut  [1]{\csname bibitem#1\endcsname}%
\let\auto@bib@innerbib\@empty
\bibitem [{\citenamefont {Jiang}\ \emph {et~al.}(2011)\citenamefont {Jiang},
  \citenamefont {Jin},\ and\ \citenamefont {Wu}}]{Jiang2011}%
  \BibitemOpen
  \bibfield  {author} {\bibinfo {author} {\bibfnamefont {D.-e.}\ \bibnamefont
  {Jiang}}, \bibinfo {author} {\bibfnamefont {Z.}~\bibnamefont {Jin}}, \ and\
  \bibinfo {author} {\bibfnamefont {J.}~\bibnamefont {Wu}},\ }\href
  {http://dx.doi.org/10.1021/nl202952d} {\bibfield  {journal} {\bibinfo
  {journal} {Nano Lett.}\ }\textbf {\bibinfo {volume} {11}},\ \bibinfo {pages}
  {5373} (\bibinfo {year} {2011})}\BibitemShut {NoStop}%
\bibitem [{\citenamefont {Kornyshev}(2013)}]{Kornyshev2013}%
  \BibitemOpen
  \bibfield  {author} {\bibinfo {author} {\bibfnamefont {A.~A.}\ \bibnamefont
  {Kornyshev}},\ }\href {\doibase 10.1039/C3FD00026E} {\bibfield  {journal}
  {\bibinfo  {journal} {Faraday Discuss.}\ }\textbf {\bibinfo {volume} {164}},\
  \bibinfo {pages} {117} (\bibinfo {year} {2013})}\BibitemShut {NoStop}%
\bibitem [{\citenamefont {Rochester}\ \emph {et~al.}(2016)\citenamefont
  {Rochester}, \citenamefont {Kondrat}, \citenamefont {Pruessner},\ and\
  \citenamefont {Kornyshev}}]{Rochester2016}%
  \BibitemOpen
  \bibfield  {author} {\bibinfo {author} {\bibfnamefont {C.~C.}\ \bibnamefont
  {Rochester}}, \bibinfo {author} {\bibfnamefont {S.}~\bibnamefont {Kondrat}},
  \bibinfo {author} {\bibfnamefont {G.}~\bibnamefont {Pruessner}}, \ and\
  \bibinfo {author} {\bibfnamefont {A.~A.}\ \bibnamefont {Kornyshev}},\ }\href
  {http://dx.doi.org/10.1021/acs.jpcc.5b12730} {\bibfield  {journal} {\bibinfo
  {journal} {J. Phys. Chem. C}\ }\textbf {\bibinfo {volume} {120}},\ \bibinfo
  {pages} {16042} (\bibinfo {year} {2016})}\BibitemShut {NoStop}%
\bibitem [{\citenamefont {Kong}\ \emph {et~al.}(2015)\citenamefont {Kong},
  \citenamefont {Wu},\ and\ \citenamefont {Henderson}}]{Kong2015}%
  \BibitemOpen
  \bibfield  {author} {\bibinfo {author} {\bibfnamefont {X.}~\bibnamefont
  {Kong}}, \bibinfo {author} {\bibfnamefont {J.}~\bibnamefont {Wu}}, \ and\
  \bibinfo {author} {\bibfnamefont {D.}~\bibnamefont {Henderson}},\ }\href
  {http://www.sciencedirect.com/science/article/pii/S0021979714008601}
  {\bibfield  {journal} {\bibinfo  {journal} {J. Colloid Interface Sci.}\
  }\textbf {\bibinfo {volume} {449}},\ \bibinfo {pages} {130 } (\bibinfo {year}
  {2015})}\BibitemShut {NoStop}%
\bibitem [{\citenamefont {Leunissen}\ \emph {et~al.}(2005)\citenamefont
  {Leunissen}, \citenamefont {Christova}, \citenamefont {Hynninen},
  \citenamefont {Royall}, \citenamefont {Campbell}, \citenamefont {Imhof},
  \citenamefont {Dijkstra}, \citenamefont {van Roij},\ and\ \citenamefont {van
  Blaaderen}}]{Leunissen2005}%
  \BibitemOpen
  \bibfield  {author} {\bibinfo {author} {\bibfnamefont {M.~E.}\ \bibnamefont
  {Leunissen}}, \bibinfo {author} {\bibfnamefont {C.~G.}\ \bibnamefont
  {Christova}}, \bibinfo {author} {\bibfnamefont {A.-P.}\ \bibnamefont
  {Hynninen}}, \bibinfo {author} {\bibfnamefont {C.~P.}\ \bibnamefont
  {Royall}}, \bibinfo {author} {\bibfnamefont {A.~I.}\ \bibnamefont
  {Campbell}}, \bibinfo {author} {\bibfnamefont {A.}~\bibnamefont {Imhof}},
  \bibinfo {author} {\bibfnamefont {M.}~\bibnamefont {Dijkstra}}, \bibinfo
  {author} {\bibfnamefont {R.}~\bibnamefont {van Roij}}, \ and\ \bibinfo
  {author} {\bibfnamefont {A.}~\bibnamefont {van Blaaderen}},\ }\href
  {http://dx.doi.org/10.1038/nature03946} {\bibfield  {journal} {\bibinfo
  {journal} {Nature}\ }\textbf {\bibinfo {volume} {437}},\ \bibinfo {pages}
  {235 } (\bibinfo {year} {2005})}\BibitemShut {NoStop}%
\bibitem [{\citenamefont {Demir\"ors}\ \emph {et~al.}(2015)\citenamefont
  {Demir\"ors}, \citenamefont {Stiefelhagen}, \citenamefont {Vissers},
  \citenamefont {Smallenburg}, \citenamefont {Dijkstra}, \citenamefont
  {Imhof},\ and\ \citenamefont {van Blaaderen}}]{Demirors2015}%
  \BibitemOpen
  \bibfield  {author} {\bibinfo {author} {\bibfnamefont {A.~F.}\ \bibnamefont
  {Demir\"ors}}, \bibinfo {author} {\bibfnamefont {J.~C.~P.}\ \bibnamefont
  {Stiefelhagen}}, \bibinfo {author} {\bibfnamefont {T.}~\bibnamefont
  {Vissers}}, \bibinfo {author} {\bibfnamefont {F.}~\bibnamefont
  {Smallenburg}}, \bibinfo {author} {\bibfnamefont {M.}~\bibnamefont
  {Dijkstra}}, \bibinfo {author} {\bibfnamefont {A.}~\bibnamefont {Imhof}}, \
  and\ \bibinfo {author} {\bibfnamefont {A.}~\bibnamefont {van Blaaderen}},\
  }\href@noop {} {\bibfield  {journal} {\bibinfo  {journal} {Phys. Rev. X}\
  }\textbf {\bibinfo {volume} {5}},\ \bibinfo {pages} {021012} (\bibinfo {year}
  {2015})}\BibitemShut {NoStop}%
\bibitem [{\citenamefont {W\"urger}(2008)}]{Wurger2008}%
  \BibitemOpen
  \bibfield  {author} {\bibinfo {author} {\bibfnamefont {A.}~\bibnamefont
  {W\"urger}},\ }\href@noop {} {\bibfield  {journal} {\bibinfo  {journal}
  {Phys. Rev. Lett.}\ }\textbf {\bibinfo {volume} {101}},\ \bibinfo {pages}
  {108302} (\bibinfo {year} {2008})}\BibitemShut {NoStop}%
\bibitem [{\citenamefont {Aric{\`o}}\ \emph {et~al.}(2005)\citenamefont
  {Aric{\`o}}, \citenamefont {Bruce}, \citenamefont {Scrosati}, \citenamefont
  {Tarascon},\ and\ \citenamefont {van Schalkwijk}}]{Arico2005}%
  \BibitemOpen
  \bibfield  {author} {\bibinfo {author} {\bibfnamefont {A.~S.}\ \bibnamefont
  {Aric{\`o}}}, \bibinfo {author} {\bibfnamefont {P.}~\bibnamefont {Bruce}},
  \bibinfo {author} {\bibfnamefont {B.}~\bibnamefont {Scrosati}}, \bibinfo
  {author} {\bibfnamefont {J.-M.}\ \bibnamefont {Tarascon}}, \ and\ \bibinfo
  {author} {\bibfnamefont {W.}~\bibnamefont {van Schalkwijk}},\ }\href
  {http://dx.doi.org/10.1038/nmat1368} {\bibfield  {journal} {\bibinfo
  {journal} {Nat. Mater.}\ }\textbf {\bibinfo {volume} {4}},\ \bibinfo {pages}
  {366} (\bibinfo {year} {2005})}\BibitemShut {NoStop}%
\bibitem [{\citenamefont {Montelongo}\ \emph {et~al.}(2017)\citenamefont
  {Montelongo}, \citenamefont {Sikdar}, \citenamefont {Ma}, \citenamefont
  {McIntosh}, \citenamefont {Velleman}, \citenamefont {Kucernak}, \citenamefont
  {Edel},\ and\ \citenamefont {Kornyshev}}]{Montelongo2017}%
  \BibitemOpen
  \bibfield  {author} {\bibinfo {author} {\bibfnamefont {Y.}~\bibnamefont
  {Montelongo}}, \bibinfo {author} {\bibfnamefont {D.}~\bibnamefont {Sikdar}},
  \bibinfo {author} {\bibfnamefont {Y.}~\bibnamefont {Ma}}, \bibinfo {author}
  {\bibfnamefont {A.~J.~S.}\ \bibnamefont {McIntosh}}, \bibinfo {author}
  {\bibfnamefont {L.}~\bibnamefont {Velleman}}, \bibinfo {author}
  {\bibfnamefont {A.~R.}\ \bibnamefont {Kucernak}}, \bibinfo {author}
  {\bibfnamefont {J.~B.}\ \bibnamefont {Edel}}, \ and\ \bibinfo {author}
  {\bibfnamefont {A.~A.}\ \bibnamefont {Kornyshev}},\ }\href
  {http://dx.doi.org/10.1038/nmat4969} {\bibfield  {journal} {\bibinfo
  {journal} {Nat. Mater.}\ }\textbf {\bibinfo {volume} {16}},\ \bibinfo {pages}
  {1127} (\bibinfo {year} {2017})}\BibitemShut {NoStop}%
\bibitem [{\citenamefont {Baughman}\ \emph {et~al.}(1999)\citenamefont
  {Baughman}, \citenamefont {Cui}, \citenamefont {Zakhidov}, \citenamefont
  {Iqbal}, \citenamefont {Barisci}, \citenamefont {Spinks}, \citenamefont
  {Wallace}, \citenamefont {Mazzoldi}, \citenamefont {De~Rossi}, \citenamefont
  {Rinzler}, \citenamefont {Jaschinski}, \citenamefont {Roth},\ and\
  \citenamefont {Kertesz}}]{Baughman1999}%
  \BibitemOpen
  \bibfield  {author} {\bibinfo {author} {\bibfnamefont {R.~H.}\ \bibnamefont
  {Baughman}}, \bibinfo {author} {\bibfnamefont {C.}~\bibnamefont {Cui}},
  \bibinfo {author} {\bibfnamefont {A.~A.}\ \bibnamefont {Zakhidov}}, \bibinfo
  {author} {\bibfnamefont {Z.}~\bibnamefont {Iqbal}}, \bibinfo {author}
  {\bibfnamefont {J.~N.}\ \bibnamefont {Barisci}}, \bibinfo {author}
  {\bibfnamefont {G.~M.}\ \bibnamefont {Spinks}}, \bibinfo {author}
  {\bibfnamefont {G.~G.}\ \bibnamefont {Wallace}}, \bibinfo {author}
  {\bibfnamefont {A.}~\bibnamefont {Mazzoldi}}, \bibinfo {author}
  {\bibfnamefont {D.}~\bibnamefont {De~Rossi}}, \bibinfo {author}
  {\bibfnamefont {A.~G.}\ \bibnamefont {Rinzler}}, \bibinfo {author}
  {\bibfnamefont {O.}~\bibnamefont {Jaschinski}}, \bibinfo {author}
  {\bibfnamefont {S.}~\bibnamefont {Roth}}, \ and\ \bibinfo {author}
  {\bibfnamefont {M.}~\bibnamefont {Kertesz}},\ }\href@noop {} {\bibfield
  {journal} {\bibinfo  {journal} {Science}\ }\textbf {\bibinfo {volume}
  {284}},\ \bibinfo {pages} {1340} (\bibinfo {year} {1999})}\BibitemShut
  {NoStop}%
\bibitem [{\citenamefont {Kornyshev}\ \emph {et~al.}(2017)\citenamefont
  {Kornyshev}, \citenamefont {Twidale},\ and\ \citenamefont
  {Kolomeisky}}]{Kornyshev2017}%
  \BibitemOpen
  \bibfield  {author} {\bibinfo {author} {\bibfnamefont {A.~A.}\ \bibnamefont
  {Kornyshev}}, \bibinfo {author} {\bibfnamefont {R.~M.}\ \bibnamefont
  {Twidale}}, \ and\ \bibinfo {author} {\bibfnamefont {A.~B.}\ \bibnamefont
  {Kolomeisky}},\ }\href {\doibase 10.1021/acs.jpcc.6b11385} {\bibfield
  {journal} {\bibinfo  {journal} {J. Phys. Chem. C}\ }\textbf {\bibinfo
  {volume} {121}},\ \bibinfo {pages} {7584} (\bibinfo {year}
  {2017})}\BibitemShut {NoStop}%
\bibitem [{\citenamefont {Porada}\ \emph {et~al.}(2013)\citenamefont {Porada},
  \citenamefont {Zhao}, \citenamefont {van~der Wal}, \citenamefont {Presser},\
  and\ \citenamefont {Biesheuvel}}]{Porada2013}%
  \BibitemOpen
  \bibfield  {author} {\bibinfo {author} {\bibfnamefont {S.}~\bibnamefont
  {Porada}}, \bibinfo {author} {\bibfnamefont {R.}~\bibnamefont {Zhao}},
  \bibinfo {author} {\bibfnamefont {A.}~\bibnamefont {van~der Wal}}, \bibinfo
  {author} {\bibfnamefont {V.}~\bibnamefont {Presser}}, \ and\ \bibinfo
  {author} {\bibfnamefont {P.}~\bibnamefont {Biesheuvel}},\ }\href@noop {}
  {\bibfield  {journal} {\bibinfo  {journal} {Prog. Mater. Sci.}\ }\textbf
  {\bibinfo {volume} {58}},\ \bibinfo {pages} {1388} (\bibinfo {year}
  {2013})}\BibitemShut {NoStop}%
\bibitem [{\citenamefont {Biesheuvel}\ and\ \citenamefont
  {Bazant}(2010)}]{Biesheuvel2010}%
  \BibitemOpen
  \bibfield  {author} {\bibinfo {author} {\bibfnamefont {P.~M.}\ \bibnamefont
  {Biesheuvel}}\ and\ \bibinfo {author} {\bibfnamefont {M.~Z.}\ \bibnamefont
  {Bazant}},\ }\href@noop {} {\bibfield  {journal} {\bibinfo  {journal} {Phys.
  Rev. E}\ }\textbf {\bibinfo {volume} {81}},\ \bibinfo {pages} {031502}
  (\bibinfo {year} {2010})}\BibitemShut {NoStop}%
\bibitem [{\citenamefont {H\"artel}\ \emph
  {et~al.}(2015{\natexlab{a}})\citenamefont {H\"artel}, \citenamefont
  {Janssen}, \citenamefont {Samin},\ and\ \citenamefont {van
  Roij}}]{Haertel2015JPCM}%
  \BibitemOpen
  \bibfield  {author} {\bibinfo {author} {\bibfnamefont {A.}~\bibnamefont
  {H\"artel}}, \bibinfo {author} {\bibfnamefont {M.}~\bibnamefont {Janssen}},
  \bibinfo {author} {\bibfnamefont {S.}~\bibnamefont {Samin}}, \ and\ \bibinfo
  {author} {\bibfnamefont {R.}~\bibnamefont {van Roij}},\ }\href
  {http://stacks.iop.org/0953-8984/27/i=19/a=194129} {\bibfield  {journal}
  {\bibinfo  {journal} {J. Phys. Condens. Matter}\ }\textbf {\bibinfo {volume}
  {27}},\ \bibinfo {pages} {194129} (\bibinfo {year}
  {2015}{\natexlab{a}})}\BibitemShut {NoStop}%
\bibitem [{\citenamefont {Evans}\ \emph {et~al.}(2016)\citenamefont {Evans},
  \citenamefont {Oettel}, \citenamefont {Roth},\ and\ \citenamefont
  {Kahl}}]{Evans2016}%
  \BibitemOpen
  \bibfield  {author} {\bibinfo {author} {\bibfnamefont {R.}~\bibnamefont
  {Evans}}, \bibinfo {author} {\bibfnamefont {M.}~\bibnamefont {Oettel}},
  \bibinfo {author} {\bibfnamefont {R.}~\bibnamefont {Roth}}, \ and\ \bibinfo
  {author} {\bibfnamefont {G.}~\bibnamefont {Kahl}},\ }\href
  {http://stacks.iop.org/0953-8984/28/i=24/a=240401} {\bibfield  {journal}
  {\bibinfo  {journal} {J. Phys. Condens. Matter}\ }\textbf {\bibinfo {volume}
  {28}},\ \bibinfo {pages} {240401} (\bibinfo {year} {2016})}\BibitemShut
  {NoStop}%
\bibitem [{\citenamefont {Archer}\ and\ \citenamefont
  {Evans}(2004)}]{Archer2004}%
  \BibitemOpen
  \bibfield  {author} {\bibinfo {author} {\bibfnamefont {A.~J.}\ \bibnamefont
  {Archer}}\ and\ \bibinfo {author} {\bibfnamefont {R.}~\bibnamefont {Evans}},\
  }\href {http://dx.doi.org/10.1063/1.1778374} {\bibfield  {journal} {\bibinfo
  {journal} {J. Chem. Phys.}\ }\textbf {\bibinfo {volume} {121}},\ \bibinfo
  {pages} {4246} (\bibinfo {year} {2004})}\BibitemShut {NoStop}%
\bibitem [{\citenamefont {Espa{\~n}ol}\ and\ \citenamefont
  {L\"owen}(2009)}]{Espanol2009}%
  \BibitemOpen
  \bibfield  {author} {\bibinfo {author} {\bibfnamefont {P.}~\bibnamefont
  {Espa{\~n}ol}}\ and\ \bibinfo {author} {\bibfnamefont {H.}~\bibnamefont
  {L\"owen}},\ }\href {http://dx.doi.org/10.1063/1.3266943} {\bibfield
  {journal} {\bibinfo  {journal} {J. Chem. Phys.}\ }\textbf {\bibinfo {volume}
  {131}},\ \bibinfo {pages} {244101} (\bibinfo {year} {2009})}\BibitemShut
  {NoStop}%
\bibitem [{\citenamefont {Marconi}\ and\ \citenamefont
  {Tarazona}(1999)}]{Marconi1999}%
  \BibitemOpen
  \bibfield  {author} {\bibinfo {author} {\bibfnamefont {U.~M.~B.}\
  \bibnamefont {Marconi}}\ and\ \bibinfo {author} {\bibfnamefont
  {P.}~\bibnamefont {Tarazona}},\ }\href@noop {} {\bibfield  {journal}
  {\bibinfo  {journal} {J. Chem. Phys.}\ }\textbf {\bibinfo {volume} {110}},\
  \bibinfo {pages} {8032} (\bibinfo {year} {1999})}\BibitemShut {NoStop}%
\bibitem [{\citenamefont {Royall}\ \emph {et~al.}(2003)\citenamefont {Royall},
  \citenamefont {Leunissen},\ and\ \citenamefont {van Blaaderen}}]{Royall}%
  \BibitemOpen
  \bibfield  {author} {\bibinfo {author} {\bibfnamefont {C.~P.}\ \bibnamefont
  {Royall}}, \bibinfo {author} {\bibfnamefont {M.~E.}\ \bibnamefont
  {Leunissen}}, \ and\ \bibinfo {author} {\bibfnamefont {A.}~\bibnamefont {van
  Blaaderen}},\ }\href {http://stacks.iop.org/0953-8984/15/i=48/a=017}
  {\bibfield  {journal} {\bibinfo  {journal} {J. Phys. Condens. Matter}\
  }\textbf {\bibinfo {volume} {15}},\ \bibinfo {pages} {S3581} (\bibinfo {year}
  {2003})}\BibitemShut {NoStop}%
\bibitem [{\citenamefont {Vissers}\ \emph
  {et~al.}(2011{\natexlab{a}})\citenamefont {Vissers}, \citenamefont {Wysocki},
  \citenamefont {Rex}, \citenamefont {Lowen}, \citenamefont {Royall},
  \citenamefont {Imhof},\ and\ \citenamefont {van Blaaderen}}]{lanes}%
  \BibitemOpen
  \bibfield  {author} {\bibinfo {author} {\bibfnamefont {T.}~\bibnamefont
  {Vissers}}, \bibinfo {author} {\bibfnamefont {A.}~\bibnamefont {Wysocki}},
  \bibinfo {author} {\bibfnamefont {M.}~\bibnamefont {Rex}}, \bibinfo {author}
  {\bibfnamefont {H.}~\bibnamefont {Lowen}}, \bibinfo {author} {\bibfnamefont
  {C.~P.}\ \bibnamefont {Royall}}, \bibinfo {author} {\bibfnamefont
  {A.}~\bibnamefont {Imhof}}, \ and\ \bibinfo {author} {\bibfnamefont
  {A.}~\bibnamefont {van Blaaderen}},\ }\href {\doibase 10.1039/C0SM01343A}
  {\bibfield  {journal} {\bibinfo  {journal} {Soft Matter}\ }\textbf {\bibinfo
  {volume} {7}},\ \bibinfo {pages} {2352} (\bibinfo {year}
  {2011}{\natexlab{a}})}\BibitemShut {NoStop}%
\bibitem [{\citenamefont {Vissers}\ \emph
  {et~al.}(2011{\natexlab{b}})\citenamefont {Vissers}, \citenamefont {van
  Blaaderen},\ and\ \citenamefont {Imhof}}]{bands}%
  \BibitemOpen
  \bibfield  {author} {\bibinfo {author} {\bibfnamefont {T.}~\bibnamefont
  {Vissers}}, \bibinfo {author} {\bibfnamefont {A.}~\bibnamefont {van
  Blaaderen}}, \ and\ \bibinfo {author} {\bibfnamefont {A.}~\bibnamefont
  {Imhof}},\ }\href@noop {} {\bibfield  {journal} {\bibinfo  {journal} {Phys.
  Rev. Lett.}\ }\textbf {\bibinfo {volume} {106}},\ \bibinfo {pages} {228303}
  (\bibinfo {year} {2011}{\natexlab{b}})}\BibitemShut {NoStop}%
\bibitem [{\citenamefont {Saunders}\ and\ \citenamefont
  {Vincent}(1999)}]{Saunders19991}%
  \BibitemOpen
  \bibfield  {author} {\bibinfo {author} {\bibfnamefont {B.~R.}\ \bibnamefont
  {Saunders}}\ and\ \bibinfo {author} {\bibfnamefont {B.}~\bibnamefont
  {Vincent}},\ }\href@noop {} {\bibfield  {journal} {\bibinfo  {journal} {Adv.
  Colloid Interface Sci.}\ }\textbf {\bibinfo {volume} {80}},\ \bibinfo {pages}
  {1} (\bibinfo {year} {1999})}\BibitemShut {NoStop}%
\bibitem [{\citenamefont {Holmqvist}\ \emph {et~al.}(2012)\citenamefont
  {Holmqvist}, \citenamefont {Mohanty}, \citenamefont {N\"agele}, \citenamefont
  {Schurtenberger},\ and\ \citenamefont {Heinen}}]{Holmqvist2012}%
  \BibitemOpen
  \bibfield  {author} {\bibinfo {author} {\bibfnamefont {P.}~\bibnamefont
  {Holmqvist}}, \bibinfo {author} {\bibfnamefont {P.~S.}\ \bibnamefont
  {Mohanty}}, \bibinfo {author} {\bibfnamefont {G.}~\bibnamefont {N\"agele}},
  \bibinfo {author} {\bibfnamefont {P.}~\bibnamefont {Schurtenberger}}, \ and\
  \bibinfo {author} {\bibfnamefont {M.}~\bibnamefont {Heinen}},\ }\href@noop {}
  {\bibfield  {journal} {\bibinfo  {journal} {Phys. Rev. Lett.}\ }\textbf
  {\bibinfo {volume} {109}},\ \bibinfo {pages} {048302} (\bibinfo {year}
  {2012})}\BibitemShut {NoStop}%
\bibitem [{\citenamefont {Lemay}\ \emph {et~al.}(2013)\citenamefont {Lemay},
  \citenamefont {Kang}, \citenamefont {Mathwig},\ and\ \citenamefont
  {Singh}}]{Lemay2013}%
  \BibitemOpen
  \bibfield  {author} {\bibinfo {author} {\bibfnamefont {S.~G.}\ \bibnamefont
  {Lemay}}, \bibinfo {author} {\bibfnamefont {S.}~\bibnamefont {Kang}},
  \bibinfo {author} {\bibfnamefont {K.}~\bibnamefont {Mathwig}}, \ and\
  \bibinfo {author} {\bibfnamefont {P.~S.}\ \bibnamefont {Singh}},\ }\href
  {\doibase 10.1021/ar300169d} {\bibfield  {journal} {\bibinfo  {journal} {Acc.
  Chem. Res.}\ }\textbf {\bibinfo {volume} {46}},\ \bibinfo {pages} {369}
  (\bibinfo {year} {2013})}\BibitemShut {NoStop}%
\bibitem [{\citenamefont {Catalano}\ and\ \citenamefont
  {Biesheuvel}(2018)}]{Biesheuvel2018}%
  \BibitemOpen
  \bibfield  {author} {\bibinfo {author} {\bibfnamefont {J.}~\bibnamefont
  {Catalano}}\ and\ \bibinfo {author} {\bibfnamefont {P.}~\bibnamefont
  {Biesheuvel}},\ }\href@noop {} {\bibfield  {journal} {\bibinfo  {journal}
  {arXiv:1805.10845}\ } (\bibinfo {year} {2018})}\BibitemShut {NoStop}%
\bibitem [{\citenamefont {{Gouy, M.}}(1910)}]{Gouy1910}%
  \BibitemOpen
  \bibfield  {author} {\bibinfo {author} {\bibnamefont {{Gouy, M.}}},\ }\href
  {\doibase 10.1051/jphystap:019100090045700} {\bibfield  {journal} {\bibinfo
  {journal} {J. Phys. Theor. Appl.}\ }\textbf {\bibinfo {volume} {9}},\
  \bibinfo {pages} {457} (\bibinfo {year} {1910})}\BibitemShut {NoStop}%
\bibitem [{\citenamefont {Chapman}(1913)}]{Chapman1913}%
  \BibitemOpen
  \bibfield  {author} {\bibinfo {author} {\bibfnamefont {D.~L.}\ \bibnamefont
  {Chapman}},\ }\href@noop {} {\bibfield  {journal} {\bibinfo  {journal}
  {Philos. Mag.}\ }\textbf {\bibinfo {volume} {25}},\ \bibinfo {pages} {475}
  (\bibinfo {year} {1913})}\BibitemShut {NoStop}%
\bibitem [{\citenamefont {Bikerman}(1942)}]{Bikerman1942}%
  \BibitemOpen
  \bibfield  {author} {\bibinfo {author} {\bibfnamefont {J.}~\bibnamefont
  {Bikerman}},\ }\href {https://doi.org/10.1080/14786444208520813} {\bibfield
  {journal} {\bibinfo  {journal} {Philos. Mag.}\ }\textbf {\bibinfo {volume}
  {33}},\ \bibinfo {pages} {384} (\bibinfo {year} {1942})}\BibitemShut
  {NoStop}%
\bibitem [{\citenamefont {Borukhov}\ \emph {et~al.}(1997)\citenamefont
  {Borukhov}, \citenamefont {Andelman},\ and\ \citenamefont
  {Orland}}]{Borukhov1997}%
  \BibitemOpen
  \bibfield  {author} {\bibinfo {author} {\bibfnamefont {I.}~\bibnamefont
  {Borukhov}}, \bibinfo {author} {\bibfnamefont {D.}~\bibnamefont {Andelman}},
  \ and\ \bibinfo {author} {\bibfnamefont {H.}~\bibnamefont {Orland}},\ }\href
  {https://link.aps.org/doi/10.1103/PhysRevLett.79.435} {\bibfield  {journal}
  {\bibinfo  {journal} {Phys. Rev. Lett.}\ }\textbf {\bibinfo {volume} {79}},\
  \bibinfo {pages} {435} (\bibinfo {year} {1997})}\BibitemShut {NoStop}%
\bibitem [{\citenamefont {Kornyshev}(2007)}]{Kornyshev2007}%
  \BibitemOpen
  \bibfield  {author} {\bibinfo {author} {\bibfnamefont {A.~A.}\ \bibnamefont
  {Kornyshev}},\ }\href {http://dx.doi.org/10.1021/jp067857o} {\bibfield
  {journal} {\bibinfo  {journal} {J. Phys. Chem. B}\ }\textbf {\bibinfo
  {volume} {111}},\ \bibinfo {pages} {5545} (\bibinfo {year}
  {2007})}\BibitemShut {NoStop}%
\bibitem [{\citenamefont {Burger}\ \emph {et~al.}(2012)\citenamefont {Burger},
  \citenamefont {Schlake},\ and\ \citenamefont {Wolfram}}]{Burger2012}%
  \BibitemOpen
  \bibfield  {author} {\bibinfo {author} {\bibfnamefont {M.}~\bibnamefont
  {Burger}}, \bibinfo {author} {\bibfnamefont {B.}~\bibnamefont {Schlake}}, \
  and\ \bibinfo {author} {\bibfnamefont {M.-T.}\ \bibnamefont {Wolfram}},\
  }\href {http://stacks.iop.org/0951-7715/25/i=4/a=961} {\bibfield  {journal}
  {\bibinfo  {journal} {Nonlinearity}\ }\textbf {\bibinfo {volume} {25}},\
  \bibinfo {pages} {961} (\bibinfo {year} {2012})}\BibitemShut {NoStop}%
\bibitem [{\citenamefont {Farjana~Siddiqua}(2017)}]{Siddiqua2017}%
  \BibitemOpen
  \bibfield  {author} {\bibinfo {author} {\bibfnamefont {S.~Z.}\ \bibnamefont
  {Farjana~Siddiqua}, \bibfnamefont {Zhongming~Wang}},\ }\href@noop {}
  {\bibfield  {journal} {\bibinfo  {journal} {arxiv:1801.00751}\ } (\bibinfo
  {year} {2017})}\BibitemShut {NoStop}%
\bibitem [{\citenamefont {Meng}\ \emph {et~al.}(2014)\citenamefont {Meng},
  \citenamefont {Zheng}, \citenamefont {Lin},\ and\ \citenamefont
  {Sushko}}]{Meng2014}%
  \BibitemOpen
  \bibfield  {author} {\bibinfo {author} {\bibfnamefont {D.}~\bibnamefont
  {Meng}}, \bibinfo {author} {\bibfnamefont {B.}~\bibnamefont {Zheng}},
  \bibinfo {author} {\bibfnamefont {G.}~\bibnamefont {Lin}}, \ and\ \bibinfo
  {author} {\bibfnamefont {M.~L.}\ \bibnamefont {Sushko}},\ }\href {\doibase
  10.4208/cicp.040913.120514a} {\bibfield  {journal} {\bibinfo  {journal}
  {Commun. Comput. Phys.}\ }\textbf {\bibinfo {volume} {16}},\ \bibinfo {pages}
  {1298} (\bibinfo {year} {2014})}\BibitemShut {NoStop}%
\bibitem [{\citenamefont {Qiao}\ \emph {et~al.}(2016)\citenamefont {Qiao},
  \citenamefont {Liu}, \citenamefont {Chen},\ and\ \citenamefont
  {Lu}}]{Qiao2016}%
  \BibitemOpen
  \bibfield  {author} {\bibinfo {author} {\bibfnamefont {Y.}~\bibnamefont
  {Qiao}}, \bibinfo {author} {\bibfnamefont {X.}~\bibnamefont {Liu}}, \bibinfo
  {author} {\bibfnamefont {M.}~\bibnamefont {Chen}}, \ and\ \bibinfo {author}
  {\bibfnamefont {B.}~\bibnamefont {Lu}},\ }\href {\doibase
  10.1007/s10955-016-1470-7} {\bibfield  {journal} {\bibinfo  {journal} {J.
  Stat. Phys.}\ }\textbf {\bibinfo {volume} {163}},\ \bibinfo {pages} {156}
  (\bibinfo {year} {2016})}\BibitemShut {NoStop}%
\bibitem [{\citenamefont {Jiang}\ \emph {et~al.}(2014)\citenamefont {Jiang},
  \citenamefont {Cao}, \citenamefont {en~Jiang},\ and\ \citenamefont
  {Wu}}]{Jiang2014}%
  \BibitemOpen
  \bibfield  {author} {\bibinfo {author} {\bibfnamefont {J.}~\bibnamefont
  {Jiang}}, \bibinfo {author} {\bibfnamefont {D.}~\bibnamefont {Cao}}, \bibinfo
  {author} {\bibfnamefont {D.}~\bibnamefont {en~Jiang}}, \ and\ \bibinfo
  {author} {\bibfnamefont {J.}~\bibnamefont {Wu}},\ }\href
  {http://stacks.iop.org/0953-8984/26/i=28/a=284102} {\bibfield  {journal}
  {\bibinfo  {journal} {J. Phys. Condens. Matter}\ }\textbf {\bibinfo {volume}
  {26}},\ \bibinfo {pages} {284102} (\bibinfo {year} {2014})}\BibitemShut
  {NoStop}%
\bibitem [{\citenamefont {Lian}\ \emph {et~al.}(2016)\citenamefont {Lian},
  \citenamefont {Zhao}, \citenamefont {Liu},\ and\ \citenamefont
  {Wu}}]{Lian2016}%
  \BibitemOpen
  \bibfield  {author} {\bibinfo {author} {\bibfnamefont {C.}~\bibnamefont
  {Lian}}, \bibinfo {author} {\bibfnamefont {S.}~\bibnamefont {Zhao}}, \bibinfo
  {author} {\bibfnamefont {H.}~\bibnamefont {Liu}}, \ and\ \bibinfo {author}
  {\bibfnamefont {J.}~\bibnamefont {Wu}},\ }\href@noop {} {\bibfield  {journal}
  {\bibinfo  {journal} {J. Chem. Phys.}\ }\textbf {\bibinfo {volume} {145}},\
  \bibinfo {pages} {204707} (\bibinfo {year} {2016})}\BibitemShut {NoStop}%
\bibitem [{\citenamefont {Gongadze}\ and\ \citenamefont
  {Igli{{\v{c}}}}(2015)}]{Gongadze2015}%
  \BibitemOpen
  \bibfield  {author} {\bibinfo {author} {\bibfnamefont {E.}~\bibnamefont
  {Gongadze}}\ and\ \bibinfo {author} {\bibfnamefont {A.}~\bibnamefont
  {Igli{{\v{c}}}}},\ }\href
  {http://www.sciencedirect.com/science/article/pii/S0013468615302449}
  {\bibfield  {journal} {\bibinfo  {journal} {Electrochim. Acta}\ }\textbf
  {\bibinfo {volume} {178}},\ \bibinfo {pages} {541} (\bibinfo {year}
  {2015})}\BibitemShut {NoStop}%
\bibitem [{\citenamefont {Beunis}\ \emph {et~al.}(2008)\citenamefont {Beunis},
  \citenamefont {Strubbe}, \citenamefont {Marescaux}, \citenamefont {Beeckman},
  \citenamefont {Neyts},\ and\ \citenamefont {Verschueren}}]{Beunis2008}%
  \BibitemOpen
  \bibfield  {author} {\bibinfo {author} {\bibfnamefont {F.}~\bibnamefont
  {Beunis}}, \bibinfo {author} {\bibfnamefont {F.}~\bibnamefont {Strubbe}},
  \bibinfo {author} {\bibfnamefont {M.}~\bibnamefont {Marescaux}}, \bibinfo
  {author} {\bibfnamefont {J.}~\bibnamefont {Beeckman}}, \bibinfo {author}
  {\bibfnamefont {K.}~\bibnamefont {Neyts}}, \ and\ \bibinfo {author}
  {\bibfnamefont {A.~R.~M.}\ \bibnamefont {Verschueren}},\ }\href
  {https://link.aps.org/doi/10.1103/PhysRevE.78.011502} {\bibfield  {journal}
  {\bibinfo  {journal} {Phys. Rev. E}\ }\textbf {\bibinfo {volume} {78}},\
  \bibinfo {pages} {011502} (\bibinfo {year} {2008})}\BibitemShut {NoStop}%
\bibitem [{\citenamefont {H\o{}jgaard~Olesen}\ \emph
  {et~al.}(2010)\citenamefont {H\o{}jgaard~Olesen}, \citenamefont {Bazant},\
  and\ \citenamefont {Bruus}}]{Bazant2010}%
  \BibitemOpen
  \bibfield  {author} {\bibinfo {author} {\bibfnamefont {L.}~\bibnamefont
  {H\o{}jgaard~Olesen}}, \bibinfo {author} {\bibfnamefont {M.~Z.}\ \bibnamefont
  {Bazant}}, \ and\ \bibinfo {author} {\bibfnamefont {H.}~\bibnamefont
  {Bruus}},\ }\href {https://link.aps.org/doi/10.1103/PhysRevE.82.011501}
  {\bibfield  {journal} {\bibinfo  {journal} {Phys. Rev. E}\ }\textbf {\bibinfo
  {volume} {82}},\ \bibinfo {pages} {011501} (\bibinfo {year}
  {2010})}\BibitemShut {NoStop}%
\bibitem [{\citenamefont {Feicht}\ \emph {et~al.}(2016)\citenamefont {Feicht},
  \citenamefont {Frankel},\ and\ \citenamefont {Khair}}]{Feicht2016}%
  \BibitemOpen
  \bibfield  {author} {\bibinfo {author} {\bibfnamefont {S.~E.}\ \bibnamefont
  {Feicht}}, \bibinfo {author} {\bibfnamefont {A.~E.}\ \bibnamefont {Frankel}},
  \ and\ \bibinfo {author} {\bibfnamefont {A.~S.}\ \bibnamefont {Khair}},\
  }\href {https://link.aps.org/doi/10.1103/PhysRevE.94.012601} {\bibfield
  {journal} {\bibinfo  {journal} {Phys. Rev. E}\ }\textbf {\bibinfo {volume}
  {94}},\ \bibinfo {pages} {012601} (\bibinfo {year} {2016})}\BibitemShut
  {NoStop}%
\bibitem [{\citenamefont {Levin}(2002)}]{Levin}%
  \BibitemOpen
  \bibfield  {author} {\bibinfo {author} {\bibfnamefont {Y.}~\bibnamefont
  {Levin}},\ }\href {http://stacks.iop.org/0034-4885/65/i=11/a=201} {\bibfield
  {journal} {\bibinfo  {journal} {Rep. Prog. Phys.}\ }\textbf {\bibinfo
  {volume} {65}},\ \bibinfo {pages} {1577} (\bibinfo {year}
  {2002})}\BibitemShut {NoStop}%
\bibitem [{\citenamefont {Fisher}\ and\ \citenamefont
  {Levin}(1993)}]{Fisher1993}%
  \BibitemOpen
  \bibfield  {author} {\bibinfo {author} {\bibfnamefont {M.~E.}\ \bibnamefont
  {Fisher}}\ and\ \bibinfo {author} {\bibfnamefont {Y.}~\bibnamefont {Levin}},\
  }\href {https://link.aps.org/doi/10.1103/PhysRevLett.71.3826} {\bibfield
  {journal} {\bibinfo  {journal} {Phys. Rev. Lett.}\ }\textbf {\bibinfo
  {volume} {71}},\ \bibinfo {pages} {3826} (\bibinfo {year}
  {1993})}\BibitemShut {NoStop}%
\bibitem [{\citenamefont {van Roij}\ and\ \citenamefont
  {Hansen}(1997)}]{vanRoij1997}%
  \BibitemOpen
  \bibfield  {author} {\bibinfo {author} {\bibfnamefont {R.}~\bibnamefont {van
  Roij}}\ and\ \bibinfo {author} {\bibfnamefont {J.-P.}\ \bibnamefont
  {Hansen}},\ }\href {https://link.aps.org/doi/10.1103/PhysRevLett.79.3082}
  {\bibfield  {journal} {\bibinfo  {journal} {Phys. Rev. Lett.}\ }\textbf
  {\bibinfo {volume} {79}},\ \bibinfo {pages} {3082} (\bibinfo {year}
  {1997})}\BibitemShut {NoStop}%
\bibitem [{\citenamefont {Gillespie}\ and\ \citenamefont
  {Eisenberg}(2001)}]{Gillespie2001}%
  \BibitemOpen
  \bibfield  {author} {\bibinfo {author} {\bibfnamefont {D.}~\bibnamefont
  {Gillespie}}\ and\ \bibinfo {author} {\bibfnamefont {R.~S.}\ \bibnamefont
  {Eisenberg}},\ }\href {https://link.aps.org/doi/10.1103/PhysRevE.63.061902}
  {\bibfield  {journal} {\bibinfo  {journal} {Phys. Rev. E}\ }\textbf {\bibinfo
  {volume} {63}},\ \bibinfo {pages} {061902} (\bibinfo {year}
  {2001})}\BibitemShut {NoStop}%
\bibitem [{\citenamefont {Kornyshev}\ and\ \citenamefont
  {Vilfan}(1995)}]{Kornyshev1995}%
  \BibitemOpen
  \bibfield  {author} {\bibinfo {author} {\bibfnamefont {A.~A.}\ \bibnamefont
  {Kornyshev}}\ and\ \bibinfo {author} {\bibfnamefont {I.}~\bibnamefont
  {Vilfan}},\ }\href
  {http://www.sciencedirect.com/science/article/pii/0013468694002642}
  {\bibfield  {journal} {\bibinfo  {journal} {Electrochim. Acta}\ }\textbf
  {\bibinfo {volume} {40}},\ \bibinfo {pages} {109} (\bibinfo {year}
  {1995})}\BibitemShut {NoStop}%
\bibitem [{\citenamefont {Kornyshev}\ and\ \citenamefont
  {Qiao}(2014)}]{Kornyshev2014}%
  \BibitemOpen
  \bibfield  {author} {\bibinfo {author} {\bibfnamefont {A.~A.}\ \bibnamefont
  {Kornyshev}}\ and\ \bibinfo {author} {\bibfnamefont {R.}~\bibnamefont
  {Qiao}},\ }\href {http://dx.doi.org/10.1021/jp5047062} {\bibfield  {journal}
  {\bibinfo  {journal} {J. Phys. Chem. C}\ }\textbf {\bibinfo {volume} {118}},\
  \bibinfo {pages} {18285} (\bibinfo {year} {2014})}\BibitemShut {NoStop}%
\bibitem [{\citenamefont {Messina}\ \emph {et~al.}(2000)\citenamefont
  {Messina}, \citenamefont {Holm},\ and\ \citenamefont {Kremer}}]{Messina2000}%
  \BibitemOpen
  \bibfield  {author} {\bibinfo {author} {\bibfnamefont {R.}~\bibnamefont
  {Messina}}, \bibinfo {author} {\bibfnamefont {C.}~\bibnamefont {Holm}}, \
  and\ \bibinfo {author} {\bibfnamefont {K.}~\bibnamefont {Kremer}},\ }\href
  {https://link.aps.org/doi/10.1103/PhysRevLett.85.872} {\bibfield  {journal}
  {\bibinfo  {journal} {Phys. Rev. Lett.}\ }\textbf {\bibinfo {volume} {85}},\
  \bibinfo {pages} {872} (\bibinfo {year} {2000})}\BibitemShut {NoStop}%
\bibitem [{\citenamefont {Lobaskin}\ \emph {et~al.}(2007)\citenamefont
  {Lobaskin}, \citenamefont {D\"unweg}, \citenamefont {Medebach}, \citenamefont
  {Palberg},\ and\ \citenamefont {Holm}}]{Lobaskin2007}%
  \BibitemOpen
  \bibfield  {author} {\bibinfo {author} {\bibfnamefont {V.}~\bibnamefont
  {Lobaskin}}, \bibinfo {author} {\bibfnamefont {B.}~\bibnamefont {D\"unweg}},
  \bibinfo {author} {\bibfnamefont {M.}~\bibnamefont {Medebach}}, \bibinfo
  {author} {\bibfnamefont {T.}~\bibnamefont {Palberg}}, \ and\ \bibinfo
  {author} {\bibfnamefont {C.}~\bibnamefont {Holm}},\ }\href
  {https://link.aps.org/doi/10.1103/PhysRevLett.98.176105} {\bibfield
  {journal} {\bibinfo  {journal} {Phys. Rev. Lett.}\ }\textbf {\bibinfo
  {volume} {98}},\ \bibinfo {pages} {176105} (\bibinfo {year}
  {2007})}\BibitemShut {NoStop}%
\bibitem [{\citenamefont {Fennell}\ \emph {et~al.}(2009)\citenamefont
  {Fennell}, \citenamefont {Bizjak}, \citenamefont {Vlachy},\ and\
  \citenamefont {Dill}}]{Fennell2009}%
  \BibitemOpen
  \bibfield  {author} {\bibinfo {author} {\bibfnamefont {C.~J.}\ \bibnamefont
  {Fennell}}, \bibinfo {author} {\bibfnamefont {A.}~\bibnamefont {Bizjak}},
  \bibinfo {author} {\bibfnamefont {V.}~\bibnamefont {Vlachy}}, \ and\ \bibinfo
  {author} {\bibfnamefont {K.~A.}\ \bibnamefont {Dill}},\ }\href
  {http://dx.doi.org/10.1021/jp809782z} {\bibfield  {journal} {\bibinfo
  {journal} {J. Phys. Chem. B}\ }\textbf {\bibinfo {volume} {113}},\ \bibinfo
  {pages} {6782} (\bibinfo {year} {2009})}\BibitemShut {NoStop}%
\bibitem [{\citenamefont {Girotto}\ \emph {et~al.}(2017)\citenamefont
  {Girotto}, \citenamefont {dos Santos},\ and\ \citenamefont
  {Levin}}]{Girotto2017}%
  \BibitemOpen
  \bibfield  {author} {\bibinfo {author} {\bibfnamefont {M.}~\bibnamefont
  {Girotto}}, \bibinfo {author} {\bibfnamefont {A.~P.}\ \bibnamefont {dos
  Santos}}, \ and\ \bibinfo {author} {\bibfnamefont {Y.}~\bibnamefont
  {Levin}},\ }\href@noop {} {\bibfield  {journal} {\bibinfo  {journal} {J.
  Chem. Phys.}\ }\textbf {\bibinfo {volume} {147}},\ \bibinfo {pages} {074109}
  (\bibinfo {year} {2017})}\BibitemShut {NoStop}%
\bibitem [{\citenamefont {Bazant}\ \emph {et~al.}(2004)\citenamefont {Bazant},
  \citenamefont {Thornton},\ and\ \citenamefont {Ajdari}}]{Bazant2004}%
  \BibitemOpen
  \bibfield  {author} {\bibinfo {author} {\bibfnamefont {M.~Z.}\ \bibnamefont
  {Bazant}}, \bibinfo {author} {\bibfnamefont {K.}~\bibnamefont {Thornton}}, \
  and\ \bibinfo {author} {\bibfnamefont {A.}~\bibnamefont {Ajdari}},\ }\href
  {https://link.aps.org/doi/10.1103/PhysRevE.70.021506} {\bibfield  {journal}
  {\bibinfo  {journal} {Phys. Rev. E}\ }\textbf {\bibinfo {volume} {70}},\
  \bibinfo {pages} {021506} (\bibinfo {year} {2004})}\BibitemShut {NoStop}%
\bibitem [{\citenamefont {Bazant}\ \emph {et~al.}(2009)\citenamefont {Bazant},
  \citenamefont {Kilic}, \citenamefont {Storey},\ and\ \citenamefont
  {Ajdari}}]{Bazant2009}%
  \BibitemOpen
  \bibfield  {author} {\bibinfo {author} {\bibfnamefont {M.~Z.}\ \bibnamefont
  {Bazant}}, \bibinfo {author} {\bibfnamefont {M.~S.}\ \bibnamefont {Kilic}},
  \bibinfo {author} {\bibfnamefont {B.~D.}\ \bibnamefont {Storey}}, \ and\
  \bibinfo {author} {\bibfnamefont {A.}~\bibnamefont {Ajdari}},\ }\href
  {http://www.sciencedirect.com/science/article/pii/S000186860900092X}
  {\bibfield  {journal} {\bibinfo  {journal} {Adv. Colloid Interface Sci.}\
  }\textbf {\bibinfo {volume} {152}},\ \bibinfo {pages} {48} (\bibinfo {year}
  {2009})}\BibitemShut {NoStop}%
\bibitem [{\citenamefont {H\"artel}\ \emph
  {et~al.}(2015{\natexlab{b}})\citenamefont {H\"artel}, \citenamefont {Kohl},\
  and\ \citenamefont {Schmiedeberg}}]{Haertel2015PRE}%
  \BibitemOpen
  \bibfield  {author} {\bibinfo {author} {\bibfnamefont {A.}~\bibnamefont
  {H\"artel}}, \bibinfo {author} {\bibfnamefont {M.}~\bibnamefont {Kohl}}, \
  and\ \bibinfo {author} {\bibfnamefont {M.}~\bibnamefont {Schmiedeberg}},\
  }\href {https://link.aps.org/doi/10.1103/PhysRevE.92.042310} {\bibfield
  {journal} {\bibinfo  {journal} {Phys. Rev. E}\ }\textbf {\bibinfo {volume}
  {92}},\ \bibinfo {pages} {042310} (\bibinfo {year}
  {2015}{\natexlab{b}})}\BibitemShut {NoStop}%
\bibitem [{\citenamefont {H\"artel}\ \emph {et~al.}(2016)\citenamefont
  {H\"artel}, \citenamefont {Samin},\ and\ \citenamefont {van
  Roij}}]{Haertel2016}%
  \BibitemOpen
  \bibfield  {author} {\bibinfo {author} {\bibfnamefont {A.}~\bibnamefont
  {H\"artel}}, \bibinfo {author} {\bibfnamefont {S.}~\bibnamefont {Samin}}, \
  and\ \bibinfo {author} {\bibfnamefont {R.}~\bibnamefont {van Roij}},\ }\href
  {http://stacks.iop.org/0953-8984/28/i=24/a=244007} {\bibfield  {journal}
  {\bibinfo  {journal} {J. Phys. Condens. Matter}\ }\textbf {\bibinfo {volume}
  {28}},\ \bibinfo {pages} {244007} (\bibinfo {year} {2016})}\BibitemShut
  {NoStop}%
\bibitem [{\citenamefont {H\"artel}(2017)}]{HaertelReview2017}%
  \BibitemOpen
  \bibfield  {author} {\bibinfo {author} {\bibfnamefont {A.}~\bibnamefont
  {H\"artel}},\ }\href {http://stacks.iop.org/0953-8984/29/i=42/a=423002}
  {\bibfield  {journal} {\bibinfo  {journal} {J. Phys. Condens. Matter}\
  }\textbf {\bibinfo {volume} {29}},\ \bibinfo {pages} {423002} (\bibinfo
  {year} {2017})}\BibitemShut {NoStop}%
\bibitem [{\citenamefont {Gillespie}(2015)}]{Gillespie2015}%
  \BibitemOpen
  \bibfield  {author} {\bibinfo {author} {\bibfnamefont {D.}~\bibnamefont
  {Gillespie}},\ }\href {\doibase 10.1007/s10404-014-1489-5} {\bibfield
  {journal} {\bibinfo  {journal} {Microfluid Nanofluid}\ }\textbf {\bibinfo
  {volume} {18}},\ \bibinfo {pages} {717} (\bibinfo {year} {2015})}\BibitemShut
  {NoStop}%
\bibitem [{\citenamefont {H\"artel}\ \emph
  {et~al.}(2015{\natexlab{c}})\citenamefont {H\"artel}, \citenamefont
  {Janssen}, \citenamefont {Weingarth}, \citenamefont {Presser},\ and\
  \citenamefont {van Roij}}]{Haertel2015EnEnvSci}%
  \BibitemOpen
  \bibfield  {author} {\bibinfo {author} {\bibfnamefont {A.}~\bibnamefont
  {H\"artel}}, \bibinfo {author} {\bibfnamefont {M.}~\bibnamefont {Janssen}},
  \bibinfo {author} {\bibfnamefont {D.}~\bibnamefont {Weingarth}}, \bibinfo
  {author} {\bibfnamefont {V.}~\bibnamefont {Presser}}, \ and\ \bibinfo
  {author} {\bibfnamefont {R.}~\bibnamefont {van Roij}},\ }\href {\doibase
  10.1039/C5EE01192B} {\bibfield  {journal} {\bibinfo  {journal} {Energy
  Environ. Sci.}\ }\textbf {\bibinfo {volume} {8}},\ \bibinfo {pages} {2396}
  (\bibinfo {year} {2015}{\natexlab{c}})}\BibitemShut {NoStop}%
\bibitem [{\citenamefont {Orkoulas}\ and\ \citenamefont
  {Panagiotopoulos}(1999)}]{Orkoulas1999}%
  \BibitemOpen
  \bibfield  {author} {\bibinfo {author} {\bibfnamefont {G.}~\bibnamefont
  {Orkoulas}}\ and\ \bibinfo {author} {\bibfnamefont {A.~Z.}\ \bibnamefont
  {Panagiotopoulos}},\ }\href {http://dx.doi.org/10.1063/1.477798} {\bibfield
  {journal} {\bibinfo  {journal} {J. Chem. Phys.}\ }\textbf {\bibinfo {volume}
  {110}},\ \bibinfo {pages} {1581} (\bibinfo {year} {1999})}\BibitemShut
  {NoStop}%
\bibitem [{\citenamefont {Luijten}\ \emph {et~al.}(2002)\citenamefont
  {Luijten}, \citenamefont {Fisher},\ and\ \citenamefont
  {Panagiotopoulos}}]{Luijten2002}%
  \BibitemOpen
  \bibfield  {author} {\bibinfo {author} {\bibfnamefont {E.}~\bibnamefont
  {Luijten}}, \bibinfo {author} {\bibfnamefont {M.~E.}\ \bibnamefont {Fisher}},
  \ and\ \bibinfo {author} {\bibfnamefont {A.~Z.}\ \bibnamefont
  {Panagiotopoulos}},\ }\href
  {https://link.aps.org/doi/10.1103/PhysRevLett.88.185701} {\bibfield
  {journal} {\bibinfo  {journal} {Phys. Rev. Lett.}\ }\textbf {\bibinfo
  {volume} {88}},\ \bibinfo {pages} {185701} (\bibinfo {year}
  {2002})}\BibitemShut {NoStop}%
\bibitem [{\citenamefont {Yan}\ and\ \citenamefont {de~Pablo}(1999)}]{Yan1999}%
  \BibitemOpen
  \bibfield  {author} {\bibinfo {author} {\bibfnamefont {Q.}~\bibnamefont
  {Yan}}\ and\ \bibinfo {author} {\bibfnamefont {J.~J.}\ \bibnamefont
  {de~Pablo}},\ }\href {http://dx.doi.org/10.1063/1.480282} {\bibfield
  {journal} {\bibinfo  {journal} {J. Chem. Phys.}\ }\textbf {\bibinfo {volume}
  {111}},\ \bibinfo {pages} {9509} (\bibinfo {year} {1999})}\BibitemShut
  {NoStop}%
\bibitem [{\citenamefont {Hynninen}\ \emph {et~al.}(2006)\citenamefont
  {Hynninen}, \citenamefont {Leunissen}, \citenamefont {van Blaaderen},\ and\
  \citenamefont {Dijkstra}}]{Hynnen2006}%
  \BibitemOpen
  \bibfield  {author} {\bibinfo {author} {\bibfnamefont {A.-P.}\ \bibnamefont
  {Hynninen}}, \bibinfo {author} {\bibfnamefont {M.~E.}\ \bibnamefont
  {Leunissen}}, \bibinfo {author} {\bibfnamefont {A.}~\bibnamefont {van
  Blaaderen}}, \ and\ \bibinfo {author} {\bibfnamefont {M.}~\bibnamefont
  {Dijkstra}},\ }\href {https://link.aps.org/doi/10.1103/PhysRevLett.96.018303}
  {\bibfield  {journal} {\bibinfo  {journal} {Phys. Rev. Lett.}\ }\textbf
  {\bibinfo {volume} {96}},\ \bibinfo {pages} {018303} (\bibinfo {year}
  {2006})}\BibitemShut {NoStop}%
\bibitem [{\citenamefont {Panagiotopoulos}(2002)}]{Panagiotopoulos2002}%
  \BibitemOpen
  \bibfield  {author} {\bibinfo {author} {\bibfnamefont {A.~Z.}\ \bibnamefont
  {Panagiotopoulos}},\ }\href {http://dx.doi.org/10.1063/1.1435571} {\bibfield
  {journal} {\bibinfo  {journal} {J. Chem. Phys.}\ }\textbf {\bibinfo {volume}
  {116}},\ \bibinfo {pages} {3007} (\bibinfo {year} {2002})}\BibitemShut
  {NoStop}%
\bibitem [{\citenamefont {Caillol}\ \emph {et~al.}(2002)\citenamefont
  {Caillol}, \citenamefont {Levesque},\ and\ \citenamefont
  {Weis}}]{Caillol2002}%
  \BibitemOpen
  \bibfield  {author} {\bibinfo {author} {\bibfnamefont {J.-M.}\ \bibnamefont
  {Caillol}}, \bibinfo {author} {\bibfnamefont {D.}~\bibnamefont {Levesque}}, \
  and\ \bibinfo {author} {\bibfnamefont {J.-J.}\ \bibnamefont {Weis}},\ }\href
  {http://dx.doi.org/10.1063/1.1480009} {\bibfield  {journal} {\bibinfo
  {journal} {J. Chem. Phys.}\ }\textbf {\bibinfo {volume} {116}},\ \bibinfo
  {pages} {10794} (\bibinfo {year} {2002})}\BibitemShut {NoStop}%
\bibitem [{\citenamefont {Valleau}\ and\ \citenamefont
  {Torrie}(1998)}]{Valleau1998}%
  \BibitemOpen
  \bibfield  {author} {\bibinfo {author} {\bibfnamefont {J.}~\bibnamefont
  {Valleau}}\ and\ \bibinfo {author} {\bibfnamefont {G.}~\bibnamefont
  {Torrie}},\ }\href {http://dx.doi.org/10.1063/1.475954} {\bibfield  {journal}
  {\bibinfo  {journal} {J. Chem. Phys.}\ }\textbf {\bibinfo {volume} {108}},\
  \bibinfo {pages} {5169} (\bibinfo {year} {1998})}\BibitemShut {NoStop}%
\bibitem [{\citenamefont {de~Carvalho}\ and\ \citenamefont
  {Evans}(1995)}]{Carvalho1995}%
  \BibitemOpen
  \bibfield  {author} {\bibinfo {author} {\bibfnamefont {R.~J. F.~L.}\
  \bibnamefont {de~Carvalho}}\ and\ \bibinfo {author} {\bibfnamefont
  {R.}~\bibnamefont {Evans}},\ }\href
  {http://stacks.iop.org/0953-8984/7/i=44/a=001} {\bibfield  {journal}
  {\bibinfo  {journal} {J. Phys. Condens. Matter}\ }\textbf {\bibinfo {volume}
  {7}},\ \bibinfo {pages} {L575} (\bibinfo {year} {1995})}\BibitemShut
  {NoStop}%
\bibitem [{\citenamefont {Smit}\ \emph {et~al.}(1996)\citenamefont {Smit},
  \citenamefont {Esselink},\ and\ \citenamefont {Frenkel}}]{Esselink1996}%
  \BibitemOpen
  \bibfield  {author} {\bibinfo {author} {\bibfnamefont {B.}~\bibnamefont
  {Smit}}, \bibinfo {author} {\bibfnamefont {K.}~\bibnamefont {Esselink}}, \
  and\ \bibinfo {author} {\bibfnamefont {D.}~\bibnamefont {Frenkel}},\ }\href
  {http://dx.doi.org/10.1080/00268979600100081} {\bibfield  {journal} {\bibinfo
   {journal} {Mol. Phys.}\ }\textbf {\bibinfo {volume} {87}},\ \bibinfo {pages}
  {159} (\bibinfo {year} {1996})}\BibitemShut {NoStop}%
\bibitem [{\citenamefont {Alts}\ \emph {et~al.}(1987)\citenamefont {Alts},
  \citenamefont {Nielaba}, \citenamefont {D'Aguanno},\ and\ \citenamefont
  {Forstmann}}]{Alts1987}%
  \BibitemOpen
  \bibfield  {author} {\bibinfo {author} {\bibfnamefont {T.}~\bibnamefont
  {Alts}}, \bibinfo {author} {\bibfnamefont {P.}~\bibnamefont {Nielaba}},
  \bibinfo {author} {\bibfnamefont {B.}~\bibnamefont {D'Aguanno}}, \ and\
  \bibinfo {author} {\bibfnamefont {F.}~\bibnamefont {Forstmann}},\ }\href
  {http://www.sciencedirect.com/science/article/pii/0301010487801362}
  {\bibfield  {journal} {\bibinfo  {journal} {Chem. Phys.}\ }\textbf {\bibinfo
  {volume} {111}},\ \bibinfo {pages} {223} (\bibinfo {year}
  {1987})}\BibitemShut {NoStop}%
\bibitem [{\citenamefont {Hansen}\ and\ \citenamefont
  {L\"owen}(2000)}]{HansenLoewen2000}%
  \BibitemOpen
  \bibfield  {author} {\bibinfo {author} {\bibfnamefont {J.-P.}\ \bibnamefont
  {Hansen}}\ and\ \bibinfo {author} {\bibfnamefont {H.}~\bibnamefont
  {L\"owen}},\ }\href {https://doi.org/10.1146/annurev.physchem.51.1.209}
  {\bibfield  {journal} {\bibinfo  {journal} {Annu. Rev. Phys. Chem.}\ }\textbf
  {\bibinfo {volume} {51}},\ \bibinfo {pages} {209} (\bibinfo {year}
  {2000})}\BibitemShut {NoStop}%
\bibitem [{\citenamefont {Evans}(1979)}]{Evans1979}%
  \BibitemOpen
  \bibfield  {author} {\bibinfo {author} {\bibfnamefont {R.}~\bibnamefont
  {Evans}},\ }\href {http://dx.doi.org/10.1080/00018737900101365} {\bibfield
  {journal} {\bibinfo  {journal} {Adv. Phys.}\ }\textbf {\bibinfo {volume}
  {28}},\ \bibinfo {pages} {143} (\bibinfo {year} {1979})}\BibitemShut
  {NoStop}%
\bibitem [{\citenamefont {Gillespie}\ \emph {et~al.}(2002)\citenamefont
  {Gillespie}, \citenamefont {Nonner},\ and\ \citenamefont
  {Eisenberg}}]{Gillespie2002}%
  \BibitemOpen
  \bibfield  {author} {\bibinfo {author} {\bibfnamefont {D.}~\bibnamefont
  {Gillespie}}, \bibinfo {author} {\bibfnamefont {W.}~\bibnamefont {Nonner}}, \
  and\ \bibinfo {author} {\bibfnamefont {R.~S.}\ \bibnamefont {Eisenberg}},\
  }\href {\doibase 10.1088/0953-8984/14/46/317} {\bibfield  {journal} {\bibinfo
   {journal} {J. Phys. Condens. Matter}\ }\textbf {\bibinfo {volume} {14}},\
  \bibinfo {pages} {12129} (\bibinfo {year} {2002})}\BibitemShut {NoStop}%
\bibitem [{\citenamefont {Gillespie}\ \emph {et~al.}(2003)\citenamefont
  {Gillespie}, \citenamefont {Nonner},\ and\ \citenamefont
  {Eisenberg}}]{Gillespie2003}%
  \BibitemOpen
  \bibfield  {author} {\bibinfo {author} {\bibfnamefont {D.}~\bibnamefont
  {Gillespie}}, \bibinfo {author} {\bibfnamefont {W.}~\bibnamefont {Nonner}}, \
  and\ \bibinfo {author} {\bibfnamefont {R.~S.}\ \bibnamefont {Eisenberg}},\
  }\href {https://link.aps.org/doi/10.1103/PhysRevE.68.031503} {\bibfield
  {journal} {\bibinfo  {journal} {Phys. Rev. E}\ }\textbf {\bibinfo {volume}
  {68}},\ \bibinfo {pages} {031503} (\bibinfo {year} {2003})}\BibitemShut
  {NoStop}%
\bibitem [{\citenamefont {Gillespie}\ \emph {et~al.}(2005)\citenamefont
  {Gillespie}, \citenamefont {Valisk\'{o}},\ and\ \citenamefont
  {Boda}}]{Gillespie2005}%
  \BibitemOpen
  \bibfield  {author} {\bibinfo {author} {\bibfnamefont {D.}~\bibnamefont
  {Gillespie}}, \bibinfo {author} {\bibfnamefont {M.}~\bibnamefont
  {Valisk\'{o}}}, \ and\ \bibinfo {author} {\bibfnamefont {D.}~\bibnamefont
  {Boda}},\ }\href {http://stacks.iop.org/0953-8984/17/i=42/a=002} {\bibfield
  {journal} {\bibinfo  {journal} {J. Phys. Condens. Matter}\ }\textbf {\bibinfo
  {volume} {17}},\ \bibinfo {pages} {6609} (\bibinfo {year}
  {2005})}\BibitemShut {NoStop}%
\bibitem [{\citenamefont {Gillespie}(2011)}]{Gillespie2011}%
  \BibitemOpen
  \bibfield  {author} {\bibinfo {author} {\bibfnamefont {D.}~\bibnamefont
  {Gillespie}},\ }\href {http://dx.doi.org/10.1021/jz2001908} {\bibfield
  {journal} {\bibinfo  {journal} {J. Phys. Chem. Lett.}\ }\textbf {\bibinfo
  {volume} {2}},\ \bibinfo {pages} {1178} (\bibinfo {year} {2011})}\BibitemShut
  {NoStop}%
\bibitem [{\citenamefont {Roth}\ and\ \citenamefont
  {Gillespie}(2005)}]{Roth2005}%
  \BibitemOpen
  \bibfield  {author} {\bibinfo {author} {\bibfnamefont {R.}~\bibnamefont
  {Roth}}\ and\ \bibinfo {author} {\bibfnamefont {D.}~\bibnamefont
  {Gillespie}},\ }\href
  {https://link.aps.org/doi/10.1103/PhysRevLett.95.247801} {\bibfield
  {journal} {\bibinfo  {journal} {Phys. Rev. Lett.}\ }\textbf {\bibinfo
  {volume} {95}},\ \bibinfo {pages} {247801} (\bibinfo {year}
  {2005})}\BibitemShut {NoStop}%
\bibitem [{\citenamefont {Roth}\ and\ \citenamefont
  {Gillespie}(2016)}]{Roth2016}%
  \BibitemOpen
  \bibfield  {author} {\bibinfo {author} {\bibfnamefont {R.}~\bibnamefont
  {Roth}}\ and\ \bibinfo {author} {\bibfnamefont {D.}~\bibnamefont
  {Gillespie}},\ }\href {http://stacks.iop.org/0953-8984/28/i=24/a=244006}
  {\bibfield  {journal} {\bibinfo  {journal} {J. Phys. Condens. Matter}\
  }\textbf {\bibinfo {volume} {28}},\ \bibinfo {pages} {244006} (\bibinfo
  {year} {2016})}\BibitemShut {NoStop}%
\bibitem [{\citenamefont {Patra}\ and\ \citenamefont
  {Ghosh}(1994{\natexlab{a}})}]{Patra1994}%
  \BibitemOpen
  \bibfield  {author} {\bibinfo {author} {\bibfnamefont {C.~N.}\ \bibnamefont
  {Patra}}\ and\ \bibinfo {author} {\bibfnamefont {S.~K.}\ \bibnamefont
  {Ghosh}},\ }\href {http://dx.doi.org/10.1063/1.467186} {\bibfield  {journal}
  {\bibinfo  {journal} {J. Chem. Phys.}\ }\textbf {\bibinfo {volume} {100}},\
  \bibinfo {pages} {5219} (\bibinfo {year} {1994}{\natexlab{a}})}\BibitemShut
  {NoStop}%
\bibitem [{\citenamefont {Patra}\ and\ \citenamefont
  {Ghosh}(1994{\natexlab{b}})}]{Patra1994n2}%
  \BibitemOpen
  \bibfield  {author} {\bibinfo {author} {\bibfnamefont {C.~N.}\ \bibnamefont
  {Patra}}\ and\ \bibinfo {author} {\bibfnamefont {S.~K.}\ \bibnamefont
  {Ghosh}},\ }\href {http://dx.doi.org/10.1063/1.467464} {\bibfield  {journal}
  {\bibinfo  {journal} {J. Chem. Phys.}\ }\textbf {\bibinfo {volume} {101}},\
  \bibinfo {pages} {4143} (\bibinfo {year} {1994}{\natexlab{b}})}\BibitemShut
  {NoStop}%
\bibitem [{\citenamefont {Patra}\ and\ \citenamefont
  {Ghosh}(1997)}]{Patra1997}%
  \BibitemOpen
  \bibfield  {author} {\bibinfo {author} {\bibfnamefont {C.~N.}\ \bibnamefont
  {Patra}}\ and\ \bibinfo {author} {\bibfnamefont {S.~K.}\ \bibnamefont
  {Ghosh}},\ }\href {http://dx.doi.org/10.1063/1.473374} {\bibfield  {journal}
  {\bibinfo  {journal} {J. Chem. Phys.}\ }\textbf {\bibinfo {volume} {106}},\
  \bibinfo {pages} {2762} (\bibinfo {year} {1997})}\BibitemShut {NoStop}%
\bibitem [{\citenamefont {Patra}\ and\ \citenamefont
  {Yethiraj}(1999)}]{Patra1999}%
  \BibitemOpen
  \bibfield  {author} {\bibinfo {author} {\bibfnamefont {C.~N.}\ \bibnamefont
  {Patra}}\ and\ \bibinfo {author} {\bibfnamefont {A.}~\bibnamefont
  {Yethiraj}},\ }\href {http://dx.doi.org/10.1021/jp991062i} {\bibfield
  {journal} {\bibinfo  {journal} {J. Phys. Chem. B}\ }\textbf {\bibinfo
  {volume} {103}},\ \bibinfo {pages} {6080} (\bibinfo {year}
  {1999})}\BibitemShut {NoStop}%
\bibitem [{\citenamefont {Kierlik}\ and\ \citenamefont
  {Rosinberg}(1991)}]{Kierlik1991}%
  \BibitemOpen
  \bibfield  {author} {\bibinfo {author} {\bibfnamefont {E.}~\bibnamefont
  {Kierlik}}\ and\ \bibinfo {author} {\bibfnamefont {M.~L.}\ \bibnamefont
  {Rosinberg}},\ }\href {https://link.aps.org/doi/10.1103/PhysRevA.44.5025}
  {\bibfield  {journal} {\bibinfo  {journal} {Phys. Rev. A}\ }\textbf {\bibinfo
  {volume} {44}},\ \bibinfo {pages} {5025} (\bibinfo {year}
  {1991})}\BibitemShut {NoStop}%
\bibitem [{\citenamefont {Tarazona}(2000)}]{Tarazona2000}%
  \BibitemOpen
  \bibfield  {author} {\bibinfo {author} {\bibfnamefont {P.}~\bibnamefont
  {Tarazona}},\ }\href {https://link.aps.org/doi/10.1103/PhysRevLett.84.694}
  {\bibfield  {journal} {\bibinfo  {journal} {Phys. Rev. Lett.}\ }\textbf
  {\bibinfo {volume} {84}},\ \bibinfo {pages} {694} (\bibinfo {year}
  {2000})}\BibitemShut {NoStop}%
\bibitem [{\citenamefont {Rosenfeld}(1989)}]{Rosenfeld1989}%
  \BibitemOpen
  \bibfield  {author} {\bibinfo {author} {\bibfnamefont {Y.}~\bibnamefont
  {Rosenfeld}},\ }\href {https://link.aps.org/doi/10.1103/PhysRevLett.63.980}
  {\bibfield  {journal} {\bibinfo  {journal} {Phys. Rev. Lett.}\ }\textbf
  {\bibinfo {volume} {63}},\ \bibinfo {pages} {980} (\bibinfo {year}
  {1989})}\BibitemShut {NoStop}%
\bibitem [{\citenamefont {Roth}(2010)}]{RothReview2010}%
  \BibitemOpen
  \bibfield  {author} {\bibinfo {author} {\bibfnamefont {R.}~\bibnamefont
  {Roth}},\ }\href {\doibase 10.1088/0953-8984/22/6/063102} {\bibfield
  {journal} {\bibinfo  {journal} {J. Phys. Condens. Matter}\ }\textbf {\bibinfo
  {volume} {22}},\ \bibinfo {pages} {063102} (\bibinfo {year}
  {2010})}\BibitemShut {NoStop}%
\bibitem [{\citenamefont {H\"artel}(2013)}]{HaertelPhD}%
  \BibitemOpen
  \bibfield  {author} {\bibinfo {author} {\bibfnamefont {A.}~\bibnamefont
  {H\"artel}},\ }\href@noop {} {\emph {\bibinfo {title} {Density Functional
  Theory of Hard Colloidal Particles: from Bulk to Interfaces}}}\ (\bibinfo
  {publisher} {Aachen (Shaker)},\ \bibinfo {year} {2013})\BibitemShut {NoStop}%
\bibitem [{\citenamefont {Oettel}\ \emph {et~al.}(2010)\citenamefont {Oettel},
  \citenamefont {G\"orig}, \citenamefont {H\"artel}, \citenamefont {L\"owen},
  \citenamefont {Radu},\ and\ \citenamefont {Schilling}}]{Oettel2010}%
  \BibitemOpen
  \bibfield  {author} {\bibinfo {author} {\bibfnamefont {M.}~\bibnamefont
  {Oettel}}, \bibinfo {author} {\bibfnamefont {S.}~\bibnamefont {G\"orig}},
  \bibinfo {author} {\bibfnamefont {A.}~\bibnamefont {H\"artel}}, \bibinfo
  {author} {\bibfnamefont {H.}~\bibnamefont {L\"owen}}, \bibinfo {author}
  {\bibfnamefont {M.}~\bibnamefont {Radu}}, \ and\ \bibinfo {author}
  {\bibfnamefont {T.}~\bibnamefont {Schilling}},\ }\href
  {https://link.aps.org/doi/10.1103/PhysRevE.82.051404} {\bibfield  {journal}
  {\bibinfo  {journal} {Phys. Rev. E}\ }\textbf {\bibinfo {volume} {82}},\
  \bibinfo {pages} {051404} (\bibinfo {year} {2010})}\BibitemShut {NoStop}%
\bibitem [{\citenamefont {Chaudhuri}\ \emph {et~al.}(2017)\citenamefont
  {Chaudhuri}, \citenamefont {Allahyarov}, \citenamefont {L\"owen},
  \citenamefont {Egelhaaf},\ and\ \citenamefont {Weitz}}]{Chaudhuri2017}%
  \BibitemOpen
  \bibfield  {author} {\bibinfo {author} {\bibfnamefont {M.}~\bibnamefont
  {Chaudhuri}}, \bibinfo {author} {\bibfnamefont {E.}~\bibnamefont
  {Allahyarov}}, \bibinfo {author} {\bibfnamefont {H.}~\bibnamefont {L\"owen}},
  \bibinfo {author} {\bibfnamefont {S.~U.}\ \bibnamefont {Egelhaaf}}, \ and\
  \bibinfo {author} {\bibfnamefont {D.~A.}\ \bibnamefont {Weitz}},\ }\href@noop
  {} {\bibfield  {journal} {\bibinfo  {journal} {Phys. Rev. Lett.}\ }\textbf
  {\bibinfo {volume} {119}},\ \bibinfo {pages} {128001} (\bibinfo {year}
  {2017})}\BibitemShut {NoStop}%
\bibitem [{\citenamefont {Oleksy}\ and\ \citenamefont
  {Hansen}(2009)}]{Oleksy2009}%
  \BibitemOpen
  \bibfield  {author} {\bibinfo {author} {\bibfnamefont {A.}~\bibnamefont
  {Oleksy}}\ and\ \bibinfo {author} {\bibfnamefont {J.-P.}\ \bibnamefont
  {Hansen}},\ }\href@noop {} {\bibfield  {journal} {\bibinfo  {journal} {Mol.
  Phys.}\ }\textbf {\bibinfo {volume} {107}},\ \bibinfo {pages} {2609}
  (\bibinfo {year} {2009})}\BibitemShut {NoStop}%
\bibitem [{\citenamefont {Medasani}\ \emph {et~al.}(2014)\citenamefont
  {Medasani}, \citenamefont {Ovanesyan}, \citenamefont {Thomas}, \citenamefont
  {Sushko},\ and\ \citenamefont {Marucho}}]{Medasani2014}%
  \BibitemOpen
  \bibfield  {author} {\bibinfo {author} {\bibfnamefont {B.}~\bibnamefont
  {Medasani}}, \bibinfo {author} {\bibfnamefont {Z.}~\bibnamefont {Ovanesyan}},
  \bibinfo {author} {\bibfnamefont {D.~G.}\ \bibnamefont {Thomas}}, \bibinfo
  {author} {\bibfnamefont {M.~L.}\ \bibnamefont {Sushko}}, \ and\ \bibinfo
  {author} {\bibfnamefont {M.}~\bibnamefont {Marucho}},\ }\href@noop {}
  {\bibfield  {journal} {\bibinfo  {journal} {J. Chem. Phys.}\ }\textbf
  {\bibinfo {volume} {140}},\ \bibinfo {pages} {204510} (\bibinfo {year}
  {2014})}\BibitemShut {NoStop}%
\end{thebibliography}%

\end{document}


\noindent\textbf{\LARGE Supporting Information}\\[2cm]
\noindent\textbf{\Large Impedance resonance in narrow confinement}\\[0.5cm]
Sonja Babel, Institut f\"ur Theoretische Physik II: Weiche Materie, Heinrich-Heine-Universit\"at D\"usseldorf, Universit\"atsstra\ss e 1, D-40225 D\"usseldorf, Germany\\ 

\noindent Michael Eikerling,
Department of Chemistry, Simon Fraser University, 8888 University Drive, Burnaby, British Columbia, Canada V5A 1S6\\ \noindent 

\noindent Hartmut L\"owen,
Institut f\"ur Theoretische Physik II: Weiche Materie, Heinrich-Heine-Universit\"at D\"usseldorf, Universit\"atsstra\ss e 1, D-40225 D\"usseldorf, Germany
\newpage 
\section{Non-interacting ions for negligible thermal motion:\\ Low and medium frequency current}
The position of the ions in the case without walls is
\begin{equation}
z(t)=z_0+\frac{qE_0}{\omega\gamma}\sin(\omega t)\,.
\end{equation}
In the low frequency limit the length that the ions travel is larger than the system length, $d_{\rm tr}=2qE_0/\omega\gamma > L$, and the effect of the walls needs to be taken into account. All ions hit the wall at some time and stay there until the field reverses. Though all ions may start at different points $z_0$, during the first oscillation all get stuck at the wall and start moving away from it at the same time, thus ending up in the same oscillation mode. The equation for this mode over one period $T=2\pi/\omega$ is given by
\begin{align}
z(t)=\begin{cases}
\frac{d_{\rm tr}}{2}\left(\sin\left(\omega t\right)+1\right) &\text{for $t_0\le t<t_1$}\\
L &\text{for $t_1\le t<t_2$}\\
L+\frac{d_{\rm tr}}{2}\left(\sin\left(\omega t\right)-1\right) &\text{for $t_2\le t<t_3$}\\
0&\text{for $t_3\le t<t_0+T$}
\end{cases}
\end{align}
with $t_0=-T/4$, $t_1=\arcsin\left(2L/d_{\rm tr}-1\right)/\omega$, $t_2=T/4$ and $t_3=t_1+T/2$. The current density in the middle of the system produced by the ions is then 
\begin{equation}
j(z=L/2,t)=q\omega L\rho_0\left[\delta\left(\omega\left(t-t'\right)\right)-\delta\left(\omega \left(t-t''\right)\right)\right]\propto\omega\label{eq_j_lowf}
\end{equation}
with the two times $t'=\left(\arcsin\left(\frac{L}{d_{\rm tr}}-1\right)\right)/{\omega}$ and $t''=t'+T/2$ at which the ions cross the centre of the system. From this we can also extract the phase of the impedance
\begin{align}
\varphi=2\pi\frac{t'}{T}=\arcsin\left(\frac{L}{d_{\rm tr}}-1\right)
\end{align}
which is limited by the two cases $\varphi=-\pi/2$ for $d_{\rm tr}\rightarrow\infty$ or $\omega\rightarrow 0$ and $\varphi=0$ for $d_{\rm tr}=L$ or $\omega=2qE_0/\gamma L$. 
The peaked current form of eq.~(\ref{eq_j_lowf}) is broadened due to diffusion and interaction but still persists, see fig.~3.


%
At medium frequencies, where $L/2\le d_{\rm tr}\le L$ and $\frac{4q E_0}{\gamma L} \ge \omega \ge  \frac{2 qE_0}{\gamma L}$, both the ions stuck at the wall and the freely oscillating ones reach the midplane of the system. The complete current is then given as a superposition of these two contributions as 
\begin{align}
j(L/2,t)&=\frac{q^2E_0}{\gamma}\rho_0\Bigg[\cos\left(\omega t\right)\Big[\theta\left(t+\tts\right)\theta\left(\tts-t\right)+\theta\left(t-\ttss\right)\theta\left(T-\ttss-t\right)\Big]\nonumber\\
&+\delta\left(\omega\left(t+\tts\right)\right)+\delta\left(\omega\left(\tts-t\right)\right)-\delta\left(\omega\left(t-\ttss\right)\right)-\delta\left(\omega\left(T-\ttss-t\right)\right)\Bigg]\label{eq_j_middlef}
\end{align} 
for $-T/4\le t<3T/4$ and the new second passing time 
$\ttss=T/2-t'$.
The current in this simplified picture is thus a cut-off (with respect to time) oscillatory part with four additional peaks. 
While the ions that oscillate freely in the middle of the system without hitting the wall constitute the oscillatory part, ions that hit the wall all cross the plane at the same time and thus appear as current peaks at the passage time $\tts$ and return time $\ttss$ or times $-\ttss$ and $-\tts$ respectively for the opposite wall. In between the passing and returning, there are no more ions left that could cross and thus there is no current.

\section{Higher harmonics}
We consider the effect of non-linearity. From the time-dependent current, see fig.~\ref{fig_current_components}, we find a considerable amount of anharmonicity especially at small frequencies. To quantify this effect, we compare the amplitude of the main signal $j_1$ with the higher frequency contributions to $j$. 
\begin{figure}[h]
\centering
\includegraphics[width=0.45\columnwidth]{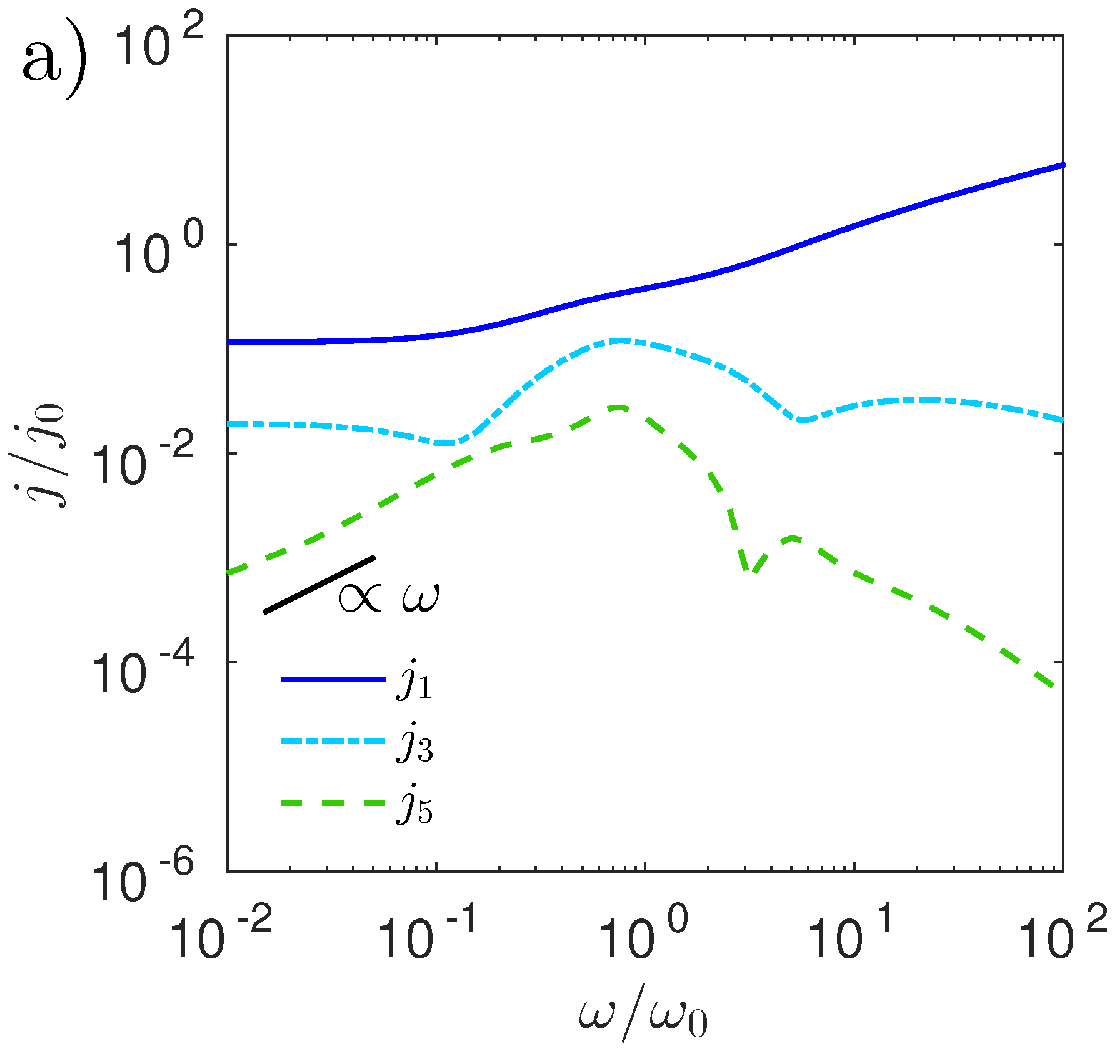}\hfill
\includegraphics[width=0.45\columnwidth]{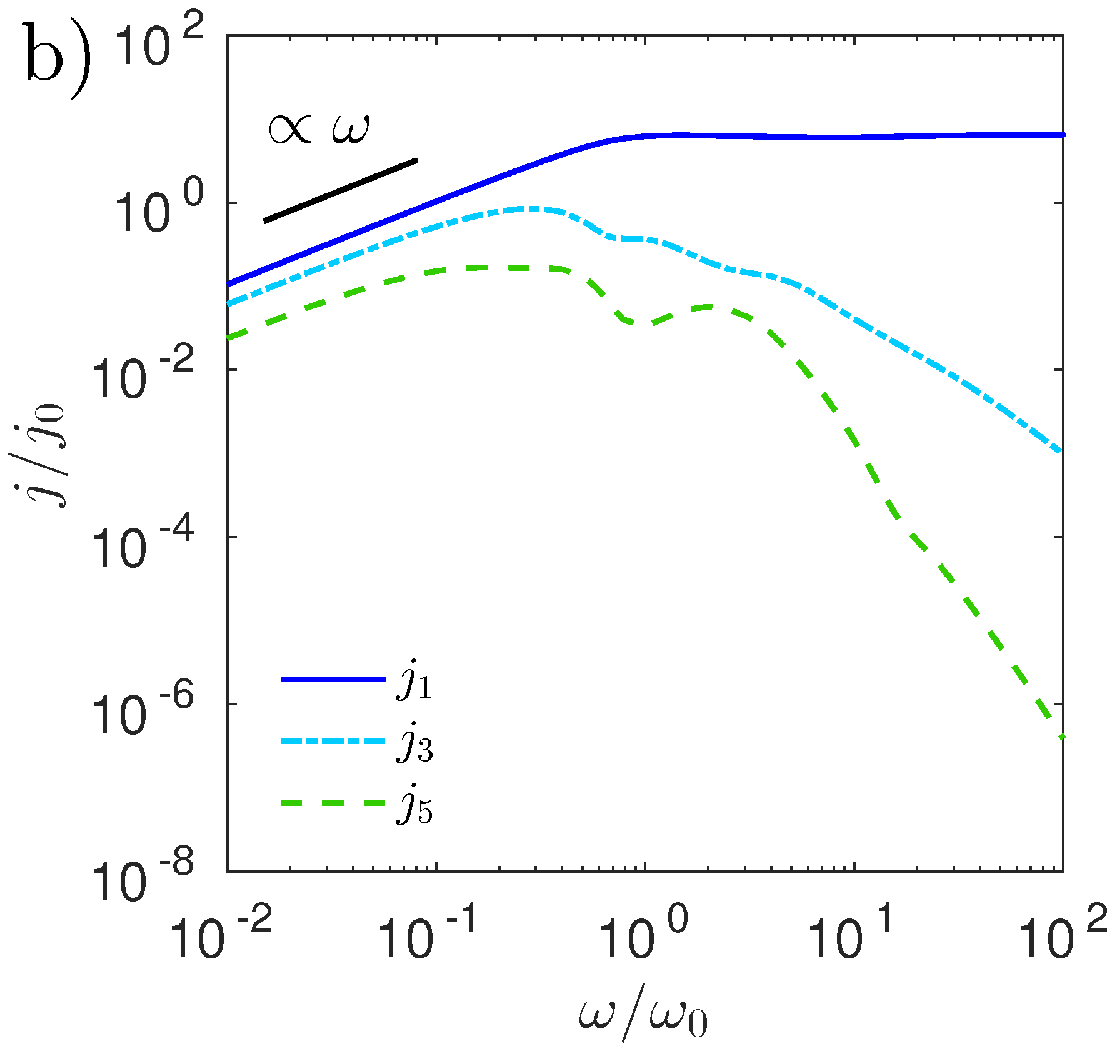}
\caption{Amplitudes of the first ($j_1$), third ($j_3$) and fifth ($j_5$) harmonic close to the wall (a) ($z/\sigma=0.01$) and at the centre (b) ($z/\sigma=2.$). While at the centre of the system all harmonics show the expected proportionality to the frequency for small $\omega$, close to the wall, the first two harmonics take a constant value. The parameters are $L/\sigma=4$, $U^*=38.9$.}\label{fig_higher_harmonics}
\end{figure}
From the symmetry of the system, which is composed of equal concentrations of oppositely charged identically sized particles, we find that $j(t+T/2)=-j(t)$ must hold. A direct consequence of this constraint is that only odd harmonics exist.
Fig.~\ref{fig_higher_harmonics} shows the amplitudes of the first, third and fifth harmonics.
The contribution of the first harmonic is always dominating though the third harmonic rises to as much as ten percent of its value. For high frequencies the amplitude of the higher harmonics drops to less than one percent.